\documentclass[12pt]{article}
\usepackage{pifont}
\usepackage{natbib}
\usepackage{graphicx}
\usepackage{amssymb}
\usepackage{epstopdf}
\usepackage{amsmath}
\usepackage{subfigure}
\usepackage{lscape}
\usepackage{latexsym}
\usepackage{amsmath}
\usepackage{yhmath}
\usepackage{bm}
\usepackage{epsfig}
\usepackage{fancyhdr}
\usepackage{sectsty,lscape,url}
\usepackage[hmargin=1in,vmargin=1in]{geometry}

\begin{document}
\pagestyle{empty}   

\title{EOF analysis of a time series with application to tsunami detection}
\author{Elena Tolkova}
\date{}
\maketitle
\begin{center}
University of Washington, Seattle WA 98195, USA\\
\vspace{0.5mm}
{\emph{e.tolkova@gmail.com}}\\
\end{center}
\vspace{1.5mm}

\begin{abstract}
Fragments of deep-ocean tidal records up to 3 days long belong to the same functional sub-space, regardless of the record's origin.  The tidal sub-space basis can be derived via Empirical Orthogonal Function (EOF) analysis of a tidal record of a single buoy.
Decomposition of a tsunami buoy record in a functional space of tidal EOFs presents an efficient tool for a short-term tidal forecast, as well as for an accurate tidal removal needed for early tsunami detection and quantification (Tolkova, E. 2009. Principal Component Analysis of Tsunami Buoy Record: Tide Prediction and Removal. Dyn. Atmos. Oceans, 46 (1-4): 62-82.) EOF analysis of a time series, however, assumes that the time series represents a stationary (in the weak sense) process.
In the present work, a modification of one-dimensional EOF formalism not restricted to stationary processes is introduced. With this modification, 
the EOF-based de-tiding/forecasting technique can be interpreted in terms of a signal passage through a filter bank, which is unique for the sub-space spanned by the EOFs. This interpretation helps to identify a harmonic content of a continuous process whose fragments are decomposed by given EOFs. In particular, seven EOFs and a constant function are proved to decompose 1-day-long tidal fragments at any location. Filtering by projection into a reduced sub-space of the above EOFs is capable of separating a tsunami wave from a tidal component within a few millimeter accuracy from the first minutes of the tsunami appearance on a tsunami buoy record, and is reliable in the presence of data gaps.
EOFs with $\sim$3-day duration (a reciprocal of either tidal band width) allow short-term (24.75 hour in advance) tidal predictions using the inherent structure of a tidal signal. The predictions do not require any a priori knowledge of tidal processes at a particular location, except for recent 49.5 hour long recordings at the location. 
\end{abstract}


\newpage

\pagestyle{fancy}       
\lhead{}

\section{Introduction}

The tsunami forecast system being developed at the Pacific Marine Environmental Laboratory relies on a network of DART (Deep-ocean Assessment and Reporting of Tsunamis) buoys to obtain real-time measurements of tsunami wave height \cite[]{vas1}. The accuracy of detecting tsunami waves at DART buoys largely determines the accuracy of any forecast of the future tsunami evolution. 

A tsunami wave in the open ocean is often less than a few centimeters in amplitude. DART stations are able to resolve tsunamis with amplitudes as small as 1 cm  \cite[]{hal3}. However, the tsunami signal is masked by the much more powerful tidal component, with typical amplitudes of 1 m or more. Therefore, for high-quality tsunami measurements, the low-frequency component of a DART record due to both tidal and non-tidal processes has to be removed with a precision of a few millimeters. 

There are two old, good techniques for removing a tidal component of a record: either predict tide or extract it using a digital filter. 

For predictions, the tidal motion is considered as a deterministic process governed by astronomical forcing. Known expansions of the tide-generating potential into harmonic constituents include tens to hundreds of terms at pre-determined frequencies, but with unknown 
phases and amplitudes (harmonic constants) that are site-specific \cite[]{thebook}, \cite[]{woce}.
The harmonic constants have to be determined for each and every individual buoy and are the most accurate when derived by means of a harmonic analysis of a particular buoy's record for one or more years \cite[]{foreman}. This technique allows to predict tide with a few cm (2 or higher) RMS error  \cite[]{kulikov}, \cite[]{hal2}, \cite[]{leprovost}. Otherwise, numerical tidal models can provide an accuracy of 3 to 9 cm. The common practice for obtaining those numbers, however, is to compare with data which were "low-pass filtered to remove the lower frequency ocean signal, and to retain only the diurnal and higher frequency part" \cite[]{leprovost}. 
Ocean signal, or long wave noise, is caused by sea level variations mainly due to fluctuations in atmospheric pressure. 
Kulikov et al (1983) estimate the typical amplitude of the natural long wave noise in the open ocean to be 1.5-2 cm, which adds to the residuals of tidal predictions.

Thus the tidal predictions alone do not provide the desired accuracy for tsunami detection and quantification. Moreover, 
DART buoys are re-deployed every two years and not necessarily into the same location, so there might not be enough data to obtain the best (data-derived) set of harmonic constituents for half of a buoy's life. 


Since tidal motion occurs on a scale of several hours, while a tsunami signal varies on a scale of several minutes, 
digital filtering is an efficient tool to separate the two processes \cite[]{foreman}, \cite[]{emery}. 
It is also commonly applied to residuals of de-tiding based on the approximation with harmonic constituents, to suppress the remaining low-frequency oscillations \cite[]{rabi}. 
Limitations of filtering from a perspective of tsunami forecast are discussed in \cite[]{kulikov}.
A digital filter has
difficulty detecting a tsunami at the beginning of the event when the ongoing tsunami signal on the end of a buoy's record is less than one full wave (a peak and a trough). In an event situation, this means waiting for the data to accumulate 30 min or longer after a tsunami wave reaches a buoy before the filtering can be relied on for tsunami detection. 
Needless to say, from the point of view of real-time tsunami forecasting, the first appearance of a tsunami wave peak on a DART record is when the precise measurement of the tsunami characteristics is essential. Gaps in the data (see Fig. \ref{gappy} for a record of the 01/03/2009 tsunami at DART 52403) present another
difficulty for digital fitering. In particular, a few-hour data gap is very likely to happen in a DART recording prior to a tsunami, as a buoy construction feature \cite[]{dart1}.

In \cite[]{me}, it was suggested to perform de-tiding and short-term tidal prediction using a set of basis functions derived from an ensemble of fragments of tidal records using principal component analysis (PCA), also known as Empirical Orthogonal Functions (EOFs) analysis.
It was noticed that a tidal fragment of a certain length, recorded anytime and anywhere in the deep ocean, can be decomposed with a few basic functions, once obtained from a single tidal record. These functions explain not only tidal but at least some variance due to low-frequency ocean processes.
The new technique was able to provide short-term tidal predictions using only a few day long tidal record at the location, and could accurately detect a tsunami signal located on the very edge of the record.

A number of questions, however, remained unanswered, with the major of them being: to what extent the functional sub-space for one buoy's tidal fragments is shared by tidal fragments recorded at other buoys?
Tidal motion is well explained in terms of harmonic analysis, whereas EOFs are not trivial to interpret in a frequency domain: "As far as frequency localization is concerned..., PCA provides no clear information" \cite[]{pairaud}. 
Consequently, all the quantitative estimates about the method accuracy, whether with regard to de-tiding or forecasting, were done solely by experimenting with a number of records, since the underlying physics of the tidal EOF method was not understood.

In this work, operations on tidal records in EOF space are interpreted in terms of conventional filtering (i.e., convolution with a sliding window). This interpretation helped to identify a harmonic content of a continuos process whose fragments belong to a reduced sub-space, given the sub-space basis vectors. The EOF basis sets developed in this work are proved to be universal, that is, be common for the majority of the DARTs. The EOF formalism was also modified compared to \cite[]{me}.
New examples of the method applications are presented, such as de-tiding a tidal record with gaps, and forecasting a tidal record a certain time (up to 24 hours) ahead by convolution with a specific sliding window 49.5 hours wide.

\section{EOFs of a temporal process: general approach}
\label{sec_formalism}

Classical EOF formalism applies to a multidimensional process/dataset $\zeta({\bf x},t)$, ${\bf x}$ being a vector of space variables and $t$ being time. EOFs of the process are found as eigenvectors of the dataset covariance matrix 
\begin{equation}
C_{space}(i,j)= \langle \zeta({\bf x_i},t) \cdot \zeta({\bf x_j},t) \rangle_t 
\label{CC}
\end{equation}
averaged over time. The eigenvectors are normalized and arranged in the order of decreasing corresponding eigenvalues (real and non-negative).
EOFs of a space-time physical process represent mutually orthogonal space modes where the data variance is concentrated, with the first mode being responsible for the largest part of the variance, the second for the largest part of the remaining variance, and so on. Remaining eigenvectors, associated with zero or negligible eigenvalues, will not be referred to as EOFs. Thus a complex process is reduced to several empirical space patterns (EOFs) evolving in time.  EOF analysis is commonly used to investigate multi-dimensional (space-time) processes \cite[]{kutzbach}, \cite[]{emery}, \cite[]{pairaud}. The EOF domain coincides with the space domain of the process. 

Less common in physical applications, 1D EOF formalism was adapted for filtering and predicting time series of climatological variables. The author is presently aware of two approaches to EOF-based time series analysis discussed in detail in \cite[]{vautard} and in \cite[]{kim1998}, with more application examples in \cite[]{vautard1} , \cite[]{mo}, \cite[]{kim1999}.

In \cite[]{kim1998}, EOFs of a temporal process $\zeta(k)$ are derived from a covariance
\begin{equation} 
C_{time}^1(i,j)= \langle \zeta(k+i) \cdot \zeta(k+j) \rangle_k,
\label{C1}
\end{equation}
and in \cite[]{vautard}, from a covariance
\begin{equation}
C_{time}^2(i,j)= \langle \zeta(k) \cdot \zeta(k+|i-j|) \rangle_k
\label{C2}
\end{equation}
where indexes $i,j$ vary from 1 to an arbitrary maximal lag $M$. The maximal lag determines an EOF domain in time, while the time domain of the process is infinite. 
An important difference with EOF formalism for space-time processes is that interpretation of time-domain EOF analysis  relies on the assumption of a process being stationary \cite[]{vautard}. The stationarity requirement is even more crucial for the prediction algorithm developed in \cite[]{kim1998}, which otherwise can not be implemented.  Therefore, unlike the multidimensional case, a covariance matrix (\ref{C1}) needs to be diagonal-constant (while a covariance (\ref{C2}) already is, regardless of the stationarity of an actual process). 

Real-world processes, however, which time-domain EOFs are intended to represent, are often quasi-periodical but not stationary, with longer period (such as seasonal or annual) modulation imposed over shorter period (such as daily) modulation. The presence of periods longer than a length of data (such as a tidal nodal cycle) would appear as non-stationarity, even if the process were stationary on a larger time scale. Non-periodical trends can also be present. "Non-stationarity" can also arise due to the statistical error of computing covariance from limited data. 
To overcome the impediment of non-stationarity, Vautard at al. (1992) derived an algorithm for removing trends and ultra-low frequencies in a record, to be performed prior to EOF analysis.  A modification of time-domain EOF formalism was introduced for a particular case of cyclostationary processes \cite[]{kim1996}, \cite[]{kim2000}.

This work (which continues \cite[]{me} with some modification and generalization of a technique introduced there) describes a different modification of 1D EOF technique.
EOF formalism is applied to $M$-reading long fragments of tidal records rather than to a continous time series. 
The fragments are viewed as independent realizations of a random vector in $M$-dimensional space. 
Prior to computing the ensemble covariance, 
a number of EOFs can be pre-assigned by removing from each fragment
a projection onto a sub-space $S$ spanned by the pre-selected vectors ${\bf s_k}$. For example, the sub-space $S$ can be spanned by a single vector ${\bf s_1}=(1, \dots, 1)^T/ \sqrt{M}$ (the constant vector). Then subtracting the $S$-projection would result in centering
each fragment on its average, which removes from the ensemble the original time series variability at a time scale larger than $M$. 
A vector ${\bf s_2}$, $s_2(i) \sim i-(M+1)/2$, can be used to remove trends prior to EOF analysis. 
Removing the $S$-projection from the data results in leading eigenvectors orthogonal to vectors ${\bf s_k}$, which are then included with an EOF basis to encompass original data fragments. 

In this work, only fragment average is removed. Thus a covariance matrix for an ensemble of $N$ tidal fragments 
is computed as
\begin{equation}
C_{i,j}=\sum_{k=1}^N {(\zeta(q_k+i-1)-a_k)(\zeta(q_k+j-1)-a_k)},
\label{cov}
\end{equation}
where $q_k$ is a node number where the $k$-th fragment starts, $a_k$ is a mean in the $k$-th fragment, $i,j=1, \dots, M$. The summation is done over $q_k$. It is possible for $q_k$ to take a value of every index in a record, but not necessary. 
$q_k$ is taken at random within the length of the record, skipping fragments with missing/corrupted data.

EOFs are derived as eigenvectors of ${\tilde{C}}_{i,j}=C_{i,j}+C_{M+1-i,M+1-j}$, that is, of a covariance component symmetrical with respect to its central element. ${\tilde{C}}$ is also a covariance of a joint ensemble of the 
de-meaned $N$ fragments  $\hat{\zeta}(q_k+i)$ and their mid-point mirrors $\hat{\zeta}(q_k+(M+1-i))$. 
So, instead of being expected to be diagonal-constant, a covariance matrix from which to derive EOFs is made symmetrical about its center
\begin{equation}
{\tilde{C}}_{M+1-i,M+1-j}={\tilde{C}}_{i,j} \ .
\label{C_sym}
\end{equation} 
It can be noted, that every diagonal-constant matrix is also symmetrical with respect to its center, but a center-symmetrical matrix (\ref{C_sym}) is not necessarily diagonally constant. 
Due to the symmetry (\ref{C_sym}), an eigenvector basis of $\tilde{C}$ is composed of even (that is, symmetrical with respect to a mid-point) and odd (anti-symmetrical with respect to a mid-point) vectors only. 

A number $n$ of EOFs sufficient to represent the original ensemble variance spans a n-dimensional sub-space in M-dimensional space referred to as a tidal sub-space. 
Due to the symmetry imposed on the EOFs, for every tidal fragment enclosed in the tidal sub-space, the sub-space also contains the fragment's time-mirror. The physical interpretation of the present modification of an EOF formalism is given in section \ref{sec_filts}. 

\section{EOFs of tidal fragments}

Tidal energy is concentrated in the long-period, diurnal, and semidiurnal frequency bands centered around 0, 1, and 2 cycles per day (c/d). A three cycles per day component (ter-diurnal principal lunar tide $M_3$) may be present with a fairly small (few mm) amplitude, so it is often neglected. Tidal components with four and more cycles per day are generated in shallow water due to nonlinearity, but are not expected in a deep ocean, where the tide is "remarkably linear" \cite[]{thebook}. The basic lunar tide $M_2$ (12.4206 h period) is the dominant tidal constituent, located in the semidiurnal band. The diurnal band is generally dominated by $K_1$ (23.9345 h period). The effective width for the diurnal band is from 0.8 to 1.1 c/d, for semidiurnal from 1.75 to 2.05 c/d; thus either bandwidth is 0.3 c/d \cite[]{munk}. Therefore, a tidal motion has two inherent time scales: one day is the apparent tidal quasi-period, and 3.3 days is the shortest length of a tidal record to resolve any individual constituent within either of the two major bands.

One-dimensional EOF formalism does not specify an EOF length, that is, the maximal lag in (\ref{C1}) or (\ref{C2}), or a fragment length  $M$. 
Vautard et al. (1992) suggested to select an eigenvector length $M$ (in units of time) so that
\begin{equation}
1/\nu_0 \le M \le 1/2\delta \nu \ ,
\label{M}
\end{equation}
where $\nu_0$ is a peak frequency in the signal spectrum, and $2\delta \nu$ is the peak width. Under this choice, EOF analysis generally isolates the corresponding oscillation, that is, a pair of oscillatory modes at $\nu_0$ frequency is expected among EOFs \cite[]{vautard}.
Our choice of $M$ also follows the limitation $M \le 1/2\delta \nu$, that is 3.3 days, but for another reason.
Since tides at different locations differ only in the fine structure of the tidal bands, one can expect the EOF sub-space to be common for all DART buoys (though their covariance matrices do look different) for as long as the fine structure of the tidal bands is not resolved. The commonality among different DART functional sub-spaces for 1-day and 3-day long fragments was noticed in \cite[]{me}. It was suggested that the tidal sub-space basis can be derived from any DART record, if the record is long enough and represented by a sufficient number of fragments. 

The EOF basis for 1-lunar-day-long tidal fragments is used for de-tiding, and a 3-lunar-day basis is used for predictions. 
A lunar day (24 h 50.4 min) is defined as in \cite[]{munk}, as a period of time between two consecutive passages of the Moon over the same meridian, or twice the $M_2$ constituent period. Records of primary interest for this work are DART records. In the absence of a tsunami, a DART transmits bottom pressure data at 15-min intervals. A lunar day, therefore, is best approached by 99 sampling intervals.

\begin{figure}[h]
	\resizebox{\textwidth}{!}
		{\includegraphics{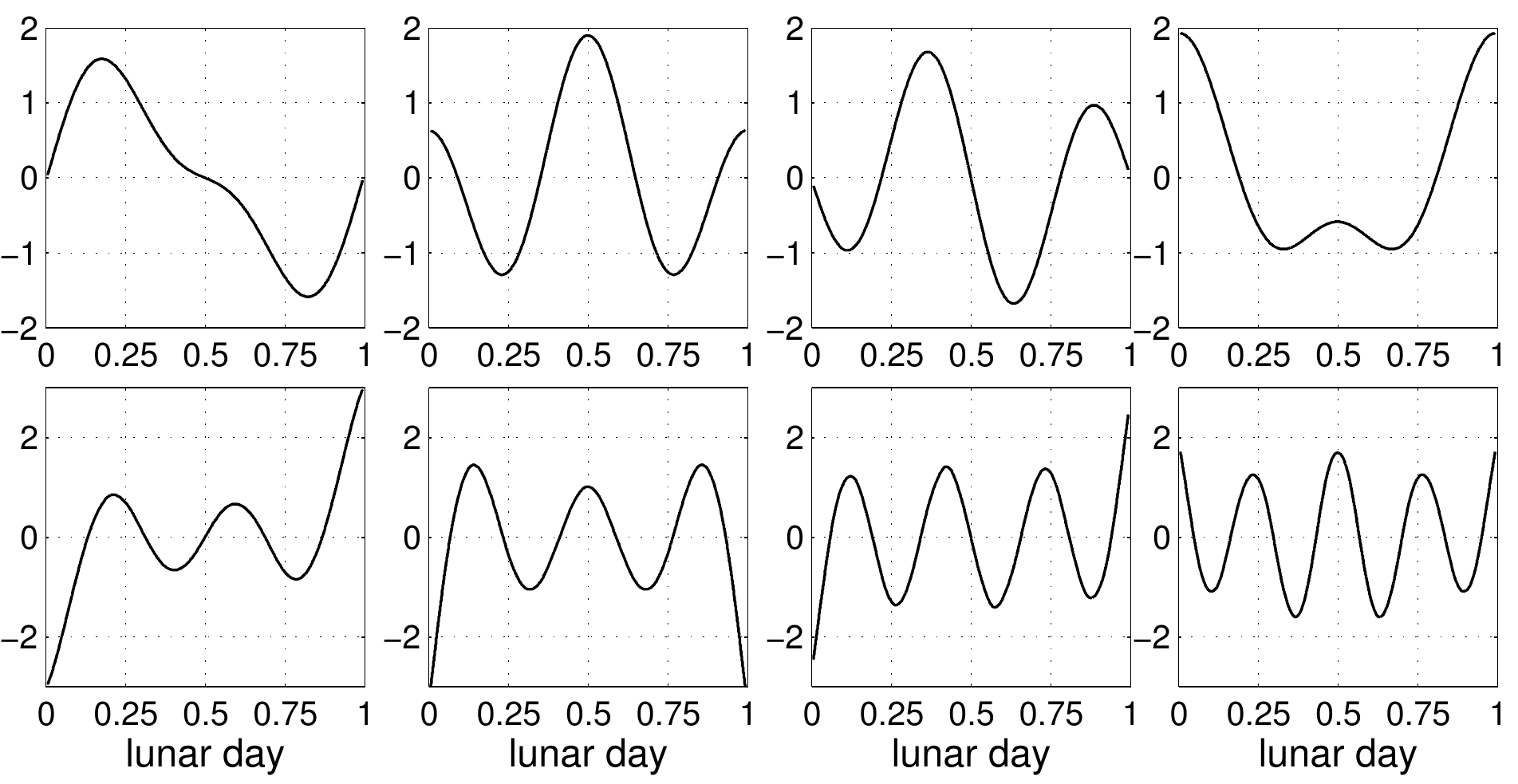}}
	\caption{
	The first eight EOFs (to be preceded by a constant function) originated with DART 46412 for 1-day-long tidal fragments. EOF amplitude is shown in proportion to the constant function amplitude.
	}
	\label{7eofs}
\end{figure} 

\begin{figure}[h]
	\centering
	\resizebox{0.8\textwidth}{!}
		{\includegraphics{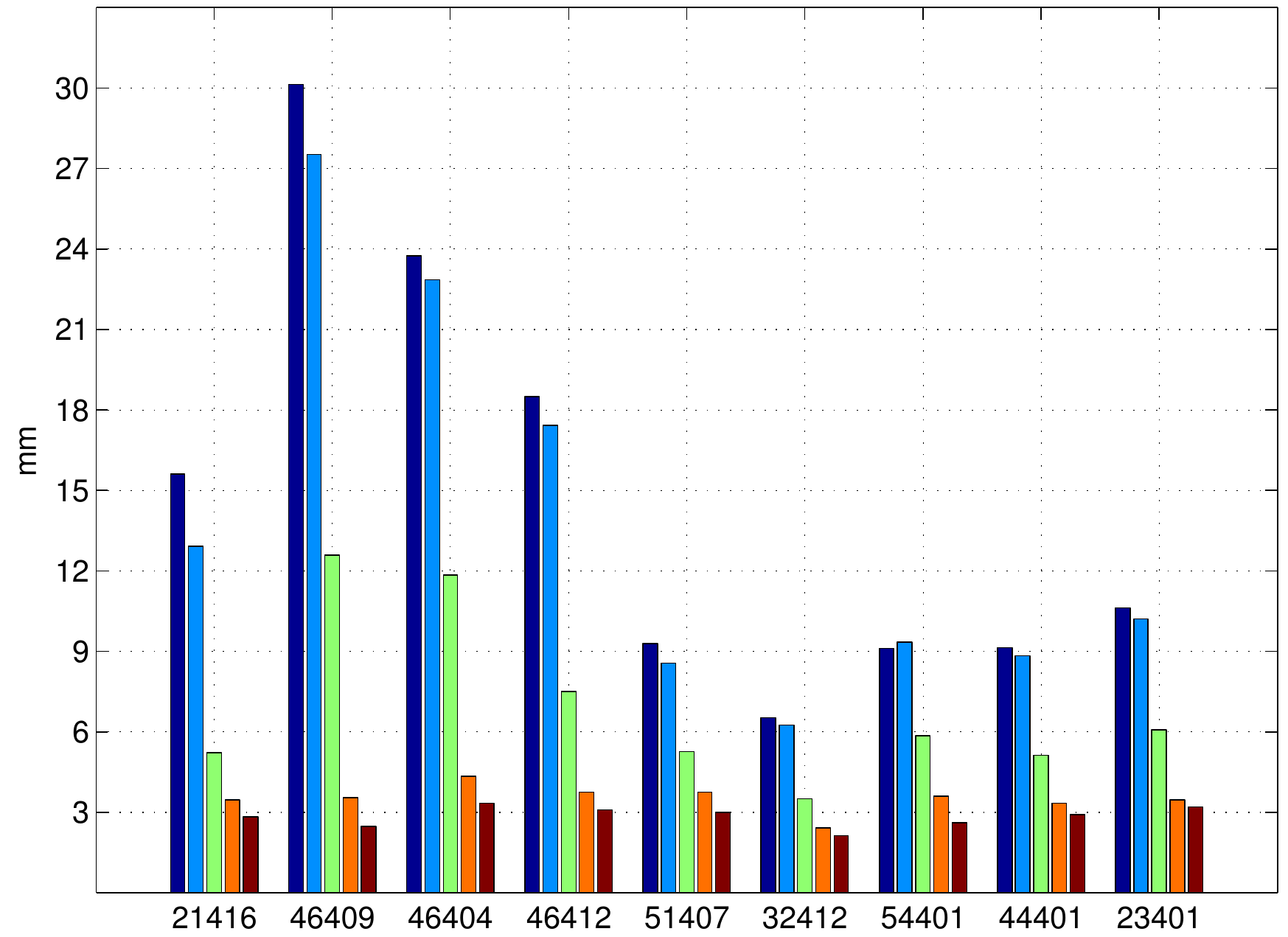}}
			\caption{
	Residuals (RMS value) of approximation of de-meaned 1-lunar-day tidal fragments with the two primary tides $K_1$ and $M_2$ (leftmost bar in each group), and with the first 4 (second left bar in each group), 5, 6, and 7 (rightmost bar in each group) tidal EOFs, in records of different DARTs.	
		}
	\label{resds}
\end{figure} 

The first eight tidal EOFs (to be preceded by the constant vector) for 1-lunar-day-long tidal fragments, computed for an ensemble of 250 fragments taken from a year of 2007 record of DART 46412, are shown in Fig. \ref{7eofs}. 
Since diurnal and semidiurnal oscillations are mutually orthogonal and (\ref{M}) is met for both, two pairs (and no more) of oscillatory modes are present.

The above EOFs are universal in the sense that they
can decompose a tidal fragment recorded anytime at any DART with the same efficiency ("signal processing" side of which is explained in section \ref{sec_filts}). 
Figure \ref{resds} shows RMS errors per reading of approximation of tidal fragments using different EOF sub-sets, and in different DART records. The DARTs were selected from every instrumented world ocean:
21416 (Pacific, Far East), 46409 (Alaska), 46404 (Pacific, North USA West Coast), 46412 (South USA West Coast), 51407 (Pacific, Hawaii), 32412 (Pacific, Southwest of Lima, Peru), 54401 (Pacific, Northeast of New Zealand), 44401 (mid-Atlantic), 23401 (Indian Ocean). 
The DART locations on a map can be found at NOAA's National Data Buoy Center public website (http://www.ndbc.noaa.gov/dart.shtml).
For each DART, the approximation errors were averaged over 70-100 tidal fragments with no missing data, randomly selected from a buoy's January-May 2008 record. For each DART,  the approximations were performed using only the first four, five, six, or seven tidal EOFs, and a constant function. A set of $n$ EOFs and a constant function will be referred to as n+1 set.
The tidal range within each buoy record was, from largest to lowest, 3.9 m for 46409; 3.5 m for 46404; 2.4 m for 46412; 1.6 m for 21416; 1.5 m for 54401; 1.0 to 1.1 m for 23401, 44401 and 51407; and 0.8 m for 32412.

An approximation with the 4+1 EOF set yielded residuals varying from 6 to 27.5 mm in RMS value (second left bar in each group in Fig. \ref{resds}). The residuals are roughly proportional to the tidal amplitude (except 54401, whose residual is level with residuals of buoys with lower tidal range). An approximation with the 7+1 EOF set (rightmost bar in each group in Fig. \ref{resds}) yields all the residuals about 3 mm and nearly the same for all the buoys, regardless of a tidal range. A DART buoy measurement sensitivity is 1 mm \cite[]{dart1}. 

For comparison with the EOF-based approximation, the leftmost bar in each group shows an average error of fitting the tidal fragments with 
a constant function, sine and cosine functions at $K_1$ primary tide frequency, and sine and cosine functions at $M_2$ primary tide frequency.  Since the constituents within each band can not be resolved with a 1-day window, fitting a 1-day fragment with a single constituent extracts most of its band variance (even if the constituent were not dominating in the band).  For the same reason, the residual is nearly orthogonal to any harmonic constituent in the band, so fitting the residual with another primary tide would not extract significant remaining variance. The residuals are generally higher than the residuals of fitting with the same number of EOFs, that is, with a 4+1 set. 
Fitting with primary tides was applied to isolate tidal variance in the 2004 Sumatra tsunami records from 200 tide stations around the world \cite[]{rabi}, \cite[]{rabithom}, \cite[]{thomrabi}.

The analysis of sub-spaces spanned by the same number of EOFs originated with different buoys showed that for the majority of the buoys,
each of the first four EOFs is buoy-specific, but altogether the first four span the same sub-space, which is roughly the sub-space of diurnal and semi-diurnal oscillations.  Each of the 5th, 6th, and 7th EOF is practically the same among different buoys, so a sub-space of the first 7 EOFs of any buoy is still common for most of the DARTs.  The 8th EOF is again buoy-specific and expands an EOF sub-space into higher frequencies. Thus different DART tidal sub-spaces intersect along 7 dimensions given by the first 7 EOFs of any buoy, though mixed-tide Alaskan and U.S. West Coast DARTs with strong tidal signals seem to be the best source for tidal EOFs.

\section{Sub-space projection as a filter bank}
\label{sec_filts}

The ability of EOFs to provide a compact representation of a signal can be used for extracting the signal component of a record by back and forth projection onto a reduced space spanned by the signal's EOF modes. 
Given a fragment of a DART record ${\bf y}$, its tidal component ${\bf z}$ is extracted by the projection onto the tidal sub-space as
\begin{equation}
{\bf z}=ff \cdot ff^T \cdot {\bf y} ,
\label{eof_filter}
\end{equation}
where ${\bf y}$ and ${\bf z}$ are M-dimensional column vectors, $ff$ is a $M \times n$ matrix with the tidal EOFs ${\bf  f}_1, \dots, {\bf f}_n$ for its columns.
Thus $M \times M$ matrix 
\begin{equation}
A=ff \cdot ff^T 
\label{mtrxA}
\end{equation}
describes an EOF filter \cite[]{raick}, \cite[]{me}. Optimally, the filter is transparent for a tidal component of a M-reading long DART record and non-transparent for a non-tidal component, such as tsunami. 

Tidal EOF sets originated with different buoys are nearly a rotation of each other in the tidal sub-space. Substituting a sub-set $ff$ for a sub-set $ff \cdot R$ rotated within the sub-space, $R$ being a $n \times n$ unitary transformation matrix ($R^T=R^{-1}$) does not change matrix A. Therefore, A is proper to the sub-space, while the sub-space basis vectors are determined up to an arbitrary unitary transformation. 
As is apparent from (\ref{mtrxA}), the sub-set basis vectors are all eigenvectors of A associated with eigenvalues equal to one, with the remaining $M-n$ eigenvalues of A being zero. Thus matrix A defines the sub-space. 

According to (\ref{eof_filter}), the $i$-th element of EOF filter output ${\bf z}$ is a weighted average of the filter input ${\bf y}$, with the weights given by the $i$-th row of $A$. Thus, every reading at the output (\ref{eof_filter}) can be interpreted as a result of a convolution of a time series with a M-reading long window, where different windows are used for different output moments, so that all the $M$ points of the output could be computed with the same M-reading long input. If, however, the input data are not limited to M readings, but continue as far as necessary into the past and future, then any single row/column of symmetrical matrix A can be used as a window function for conventional filtering via convolution with a sliding window. In particular, the last row/column of A could be a sliding window for a one-sided filter (output at every moment is computed with readings taken on or before this moment). Apparently, an EOF filter is transparent for a fragment of a process, if the process spectrum falls within a passband of each row-filter in $A$. 

An interpretation of matrix A as an ensemble of sliding windows also leads to an expectation of A having a central symmetry with respect to the matrix center: 
\begin{equation}
A(i,j)=A(M+1-i,M+1-j)
\label{A_symm}
\end{equation}
For example, the central row of A is used to compute a tidal component of a DART record on moment $t$ with the same moment reading, $(M-1)/2$ readings before and $(M-1)/2$ readings after the moment $t$. This is a reason to expect this particular window function to be symmetrical. Likewise, window functions to compute a $j$-th tidal reading from the beginning and the end of a M-reading fragment are expected to be mirrors of each other.

In terms of the sub-space basis vectors, $f_k(i)$ being the $i$-th element of the $k$-th vector, (\ref{A_symm}) can be re-written as:
\begin{equation}
\sum_{k=1}^n  {f_k(i) \cdot f_k(j)}=\sum_{k=1}^n  {f_k(M+1-i) \cdot f_k(M+1-j)}.
\label{ff_symm}
\end{equation}
The last equation would be satisfied with either odd or even (with respect to the central element) vectors ${\bf f}_k$, so the symmetry imposed on the covariance and consequently on an EOF set follows from the physical meaning of A.

\begin{figure}[h]
\centering
	\resizebox{0.9\textwidth}{!}
		{\includegraphics{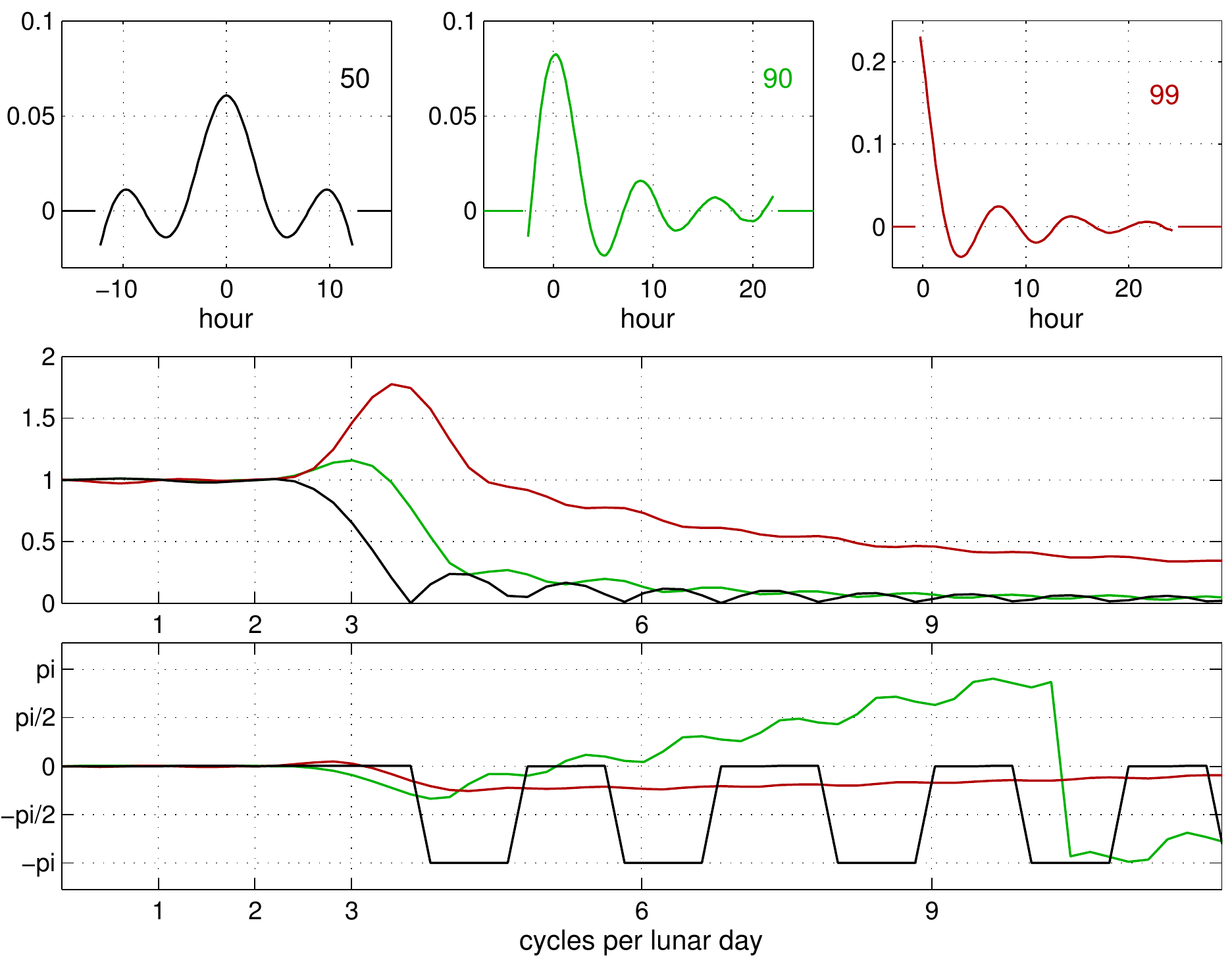}}
	\caption{
	Top panes, left to right: sliding windows in the central (50th) row, 90th row, and the last row (99th) of matrix A for 7+1 sub-set of 46412 1-lunar-day EOFs. Middle pane: amplitude characteristics of the corresponding filters in the frequency domain (the number next to each curve indicates a corresponding row number); bottom: phase characteristics of the filters.  
	}
	\label{Acolumns}
\end{figure} 

The interpretation of an EOF sub-space projection in terms of a filter bank helps to identify a harmonic content of a continuous process whose fragments belong to the EOF sub-space, given the sub-space basis vectors. Specifically, the sub-space would encompass fragments of a process whose spectrum is located within a pass band of every row-filter in an associated matrix A.

Figure \ref{Acolumns} shows pulse responses of the filters contained in the central (50th) row, 90th row, and the last (99th) row of the $99 \times 99$  matrix A computed with the 7+1 set of DART 46412 EOFs, and the amplitudes and phases of the filters' frequency responses. The three filters (and the rest of the 99) are low-pass filters which maintain high transparency and introduce no phase distortions within tidal bands. 

Therefore the above 7+1 EOF set would extract a tidal component from any 99-reading long record at any DART, as long as the tidal energy at this DART is concentrated in the three tidal bands (0, 1, and 2 c/d), that is, for all the buoys in the areas with linear astronomical tide. It also extracts low-frequency ocean noise in a band from 0 to 2 c/d. Thus the residuals of EOF de-tiding do not display trends typical for the residuals of tidal predictions. 

\begin{figure}[h]
	\centering
	\resizebox{0.8\textwidth}{!}
		{\includegraphics{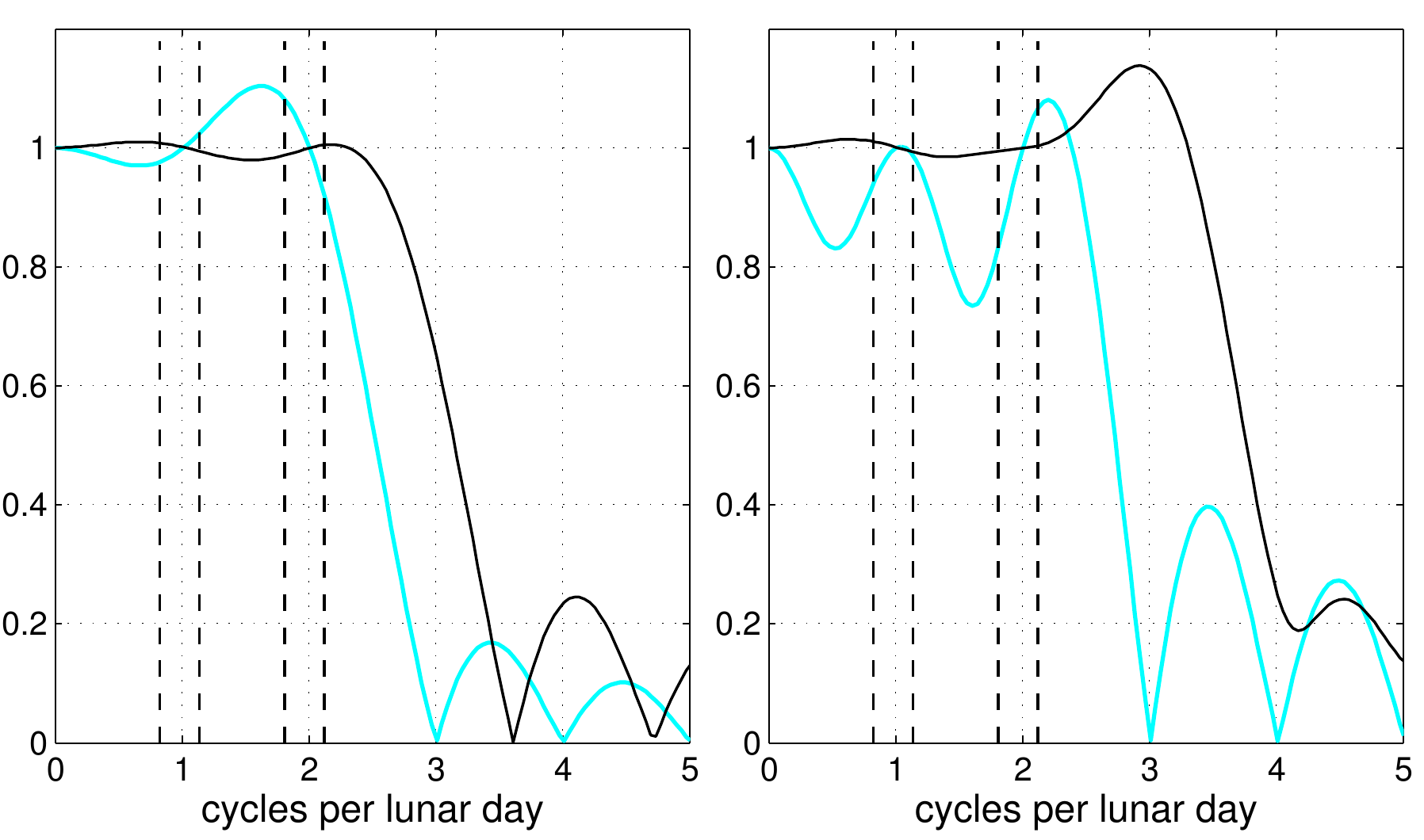}}
	\caption{
	Amplitude responses of a central row (left) and 90th row (right) filter for the 4+1 EOF set (cyan(gray)/thick) and the 7+1 set (black/thin). Dotted vertical lines denote effective width of diurnal and semidiurnal band.
	}
	\label{A4-9}
\end{figure} 

The same approach explains the relative ability of different-size EOF sub-sets to isolate tidal variance together with long wave noise within the tidal range (illustrated in Fig. \ref{resds}). 
Figure \ref{A4-9} shows amplitude responses of a central row filter and 90-th row filter for the 4+1 set and the 7+1 set of the EOFs of 46412. Note that only even EOFs contribute to the central row. The 4+1 EOF set roughly picks up the variance in the long-period, diurnal and semidiurnal bands, and so does any DART 4+1 EOF set (though the individual modes differ among DARTs). Adding the next three modes flattens the associated filter frequency response inside the bands without expanding the filter passbands significantly beyond the tidal bands. Thus the next three modes explain the remaining variance (not decomposed by the 4+1 set). 

\subsection{Effect of symmetrization}
\label{sec_symm}

\begin{figure}[h]
\centering
	\resizebox{0.9\textwidth}{!}
		{\includegraphics{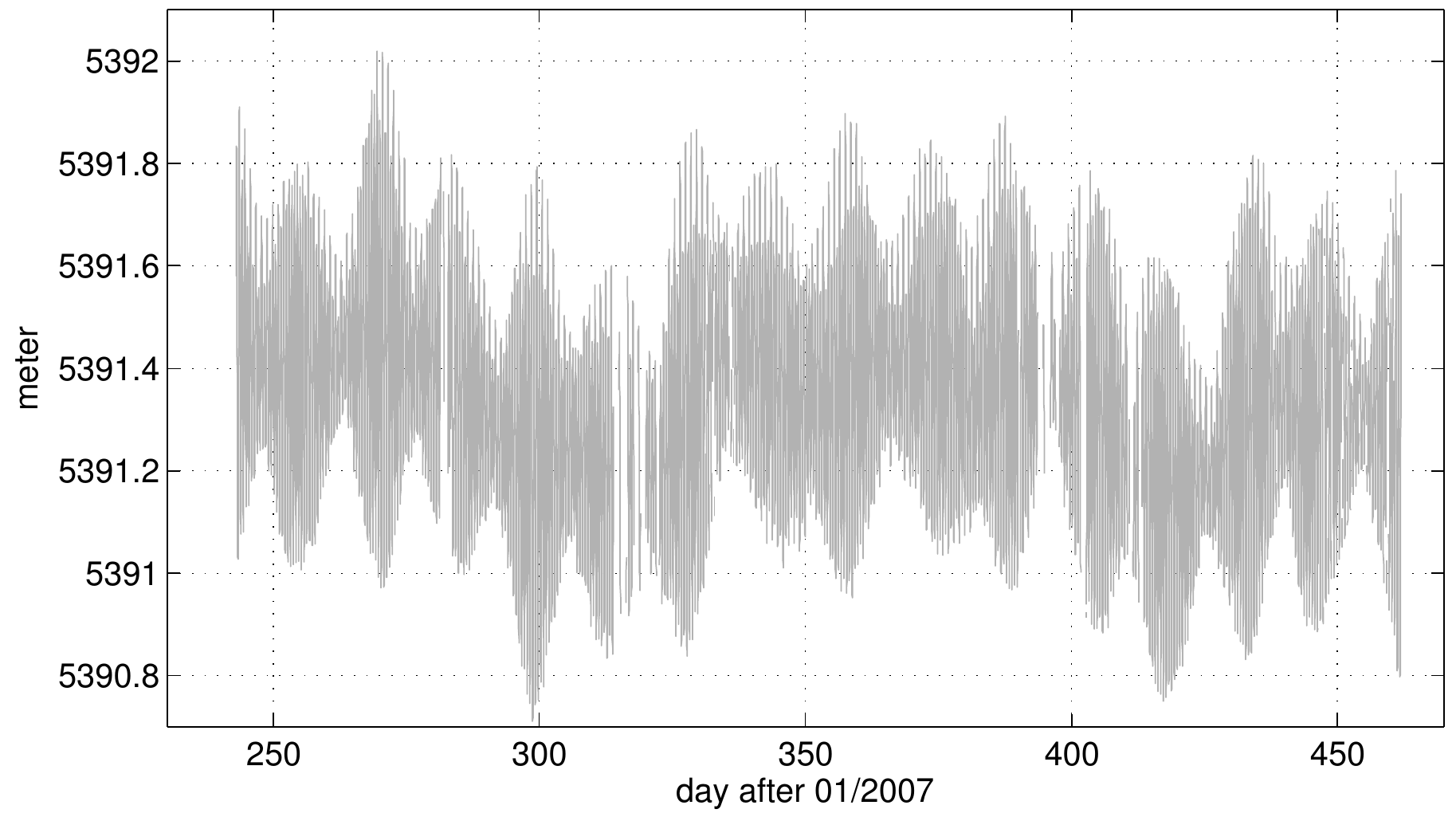}}
	\caption{
Tidal record of 44401 from 09/2007 to 03/2008.
	}
	\label{tide44401}
\end{figure} 

\begin{figure}[h]
\centering
	\resizebox{0.8\textwidth}{!}
		{\includegraphics{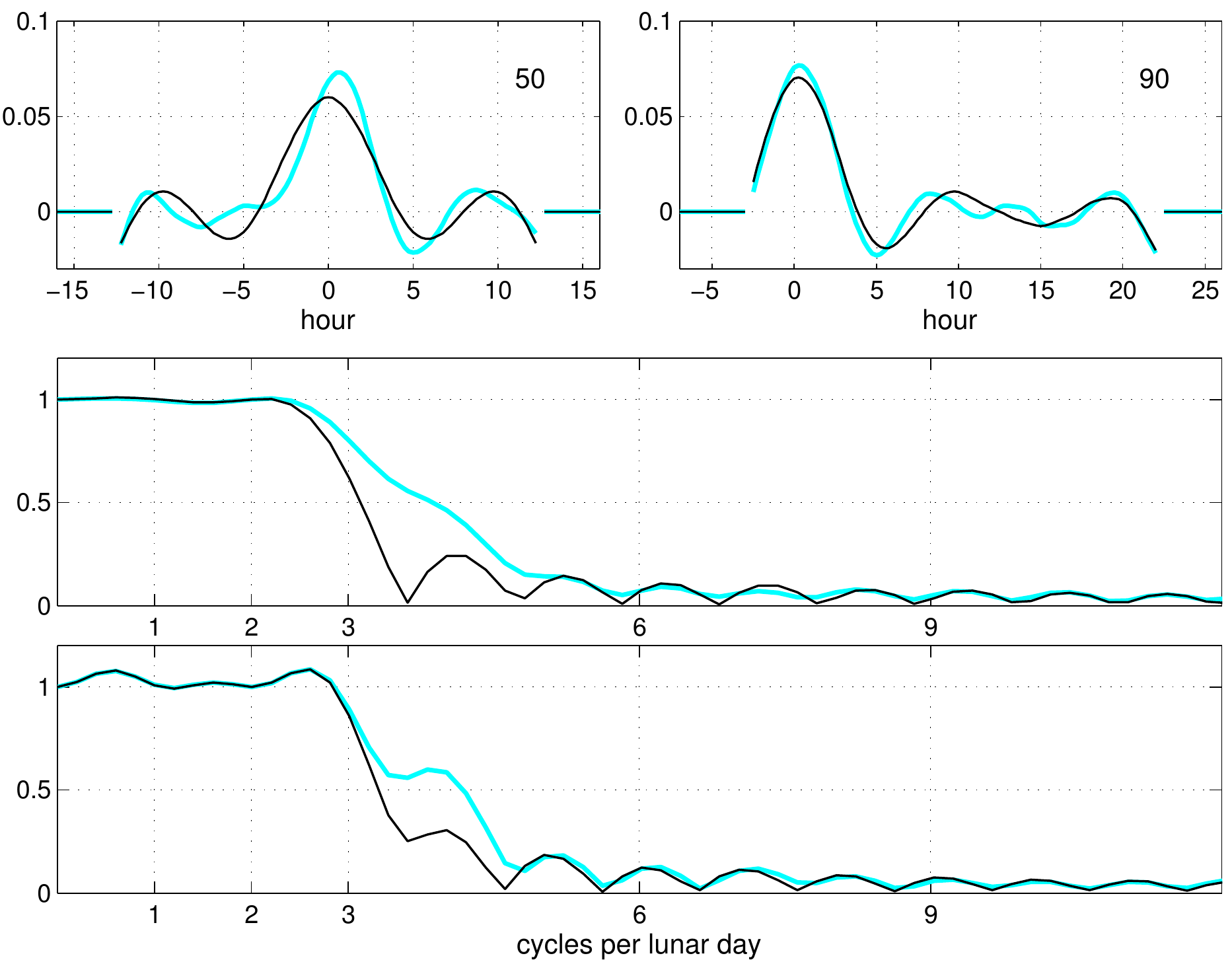}}
	\caption{
	Pulse and frequency responses of the filters in the 50th and 90th columns for a 7+1 sub-set of DART 44401, using a "true" covariance (cyan(gray)/thick) and a center-symmetrical covariance (black/thin).
	}
	\label{A44401}
\end{figure} 

An example below highlights an effect the symmetrization of a covariance has on a tidal EOF set. Two EOF sets were
computed for 130 1-day fragments randomly selected from the buoy's 09/2007-03/2008 record using a "true" covariance (\ref{cov}) and a center-symmetrical covariance $\tilde{C}$. The record has significant large-scale modulation and/or trends and contains noticeable amounts of missing data (see Figure \ref{tide44401}). 

Figure \ref{A44401} shows pulse and frequency responses of the filters in the 50th and 90th columns of matrix $A$ for each of the two 7+1 EOF sets. 
EOFs obtained with the "true" covariance are less efficient in isolating the tidal variance: the corresponding filters have passbands wider than the tidal frequency range (transparent even at 3 c/d) and ripples on the near-edge-row filter characteristics at the low frequencies (under 2 c/d).  Symmetrization reduced the passbands, but could not improve the responses of near-edge filters inside the tidal bands. Overall, filters of 44401 derived from a center-symmetrical covariance are more alike those of 46412, with the central row filters becoming practically identical.

\subsection{Passage of a wide-band signal through an EOF filter}

In the next experiment suggested by the reviewers of this paper, the spectral contribution of the EOF filtering to the residual records is estimated directly. 
A set of 13 diurnal and semidiurnal constituents, including all the major ones (M2, S2, N2, K2, K1, O1, P1, Q1) was used to model a tidal fragment with 1 min interval, which is a sampling of a DART record in an event mode. White Gaussian noise with a standard deviation $\sigma=5$ cm was added to each value in the fragment. The white noise at 1-min sampling rate occupies the entire frequency band from 0 to 30 cycles/hour. Then the fragment was EOF-filtered with the 7+1 EOF set spline-interpolated from 15 min to 1 min sampling rate. The objective was to observe the spectral effects of the EOF detiding in wide frequency range, by comparing the spectra of the original signal (a particular realization of the white noise) and the detided signal (a residual of the EOF filtering).

\begin{figure}[ht]
	\resizebox{\textwidth}{!}
		{\includegraphics{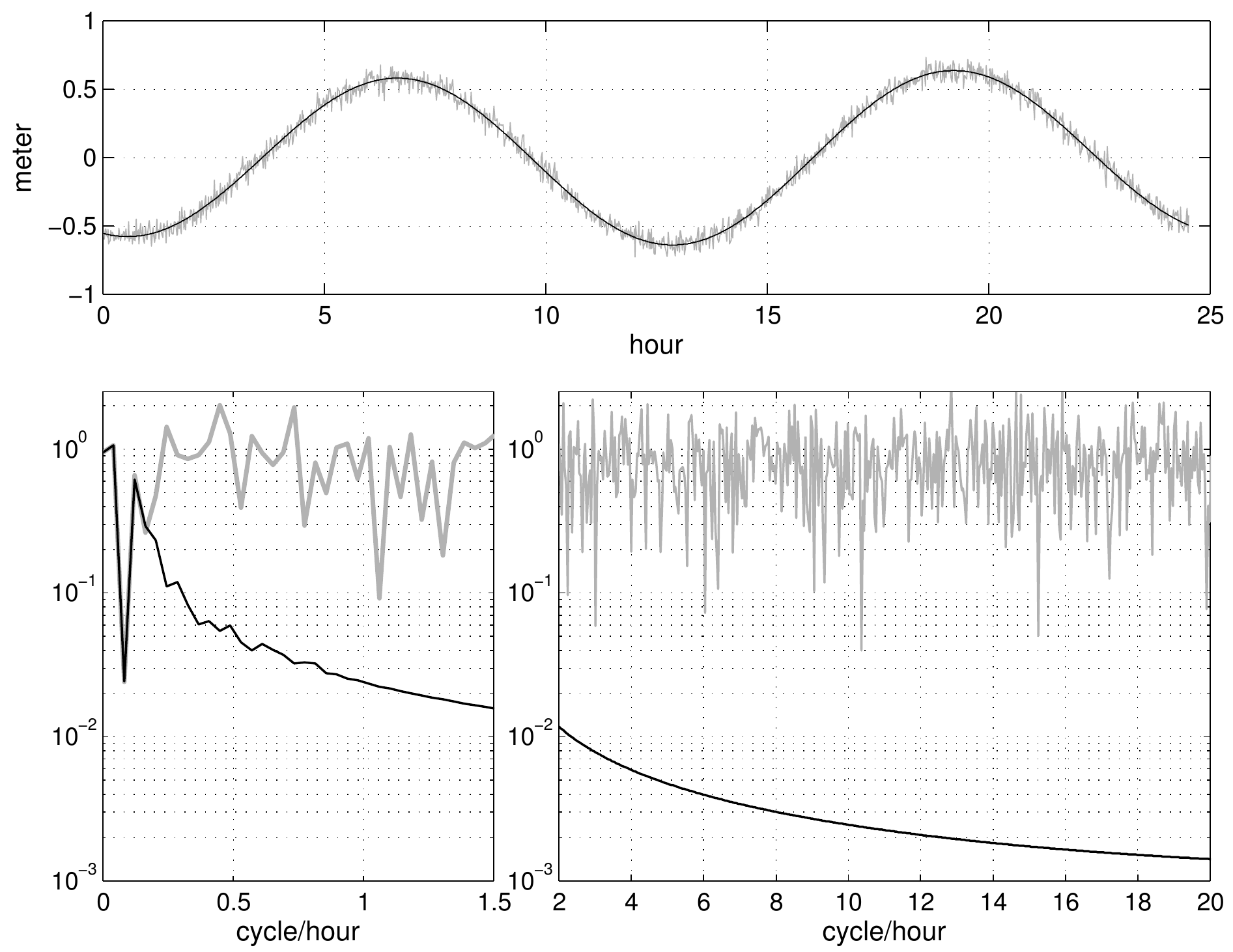}}
		\caption{Top: a tidal fragment with added white noise (gray) and a tide extracted by the EOF filter (black). Bottom: amplitude spectra of the original signal (gray) and of the difference between the original and the detided signals (black).}
		\label{wnoise}
\end{figure}

Fig. \ref{wnoise}, top pane shows a tidal fragment with added noise and its tidal component extracted by the EOF filter. Since the EOF filter extracts low frequency energy in a range from 0 to about 1/10 cycles per hour, which is 1/300 part of the entire signal band uniformly (on average) occupied by a white noise, the filter therefore captures 1/300 part of the noise energy. Thus a detiding error of about $\sigma/\sqrt{300} =2.9$ mm (RMS) per reading should be expected. An actual RMS error of the tidal estimate in the above experiment was 2.8 mm.

Fig. \ref{wnoise}, bottom panes show the amplitude spectra of the original signal (a realization of the white noise) and of the difference between the original and the detided signals. The spectral amplitude is normalized by its standard deviation, that is, $\sigma \sqrt{M}$, where $M=1471$ is a number of data points in a tidal fragment sampled at 1 min interval.
The spectral amplitude of the difference coincides with the amplitude of the original throughout the tidal bands and then monotonically decreases to 0.001 of the standard deviation of the original amplitude. Thus the residual of the EOF detiding does not contain any visible tidal energy, 
whereas at any frequency higher than 2 cycles per hour, the spectral amplitude of the residual does not differ from the original amplitude by more than 1\% .

\section{Detiding records with missing data}

So far, a tidal fragment has been considered an uninterrupted 24.75 h long recording. However, real data do have gaps. In particular, a few hour (less than 6 hour) data gap is very likely to happen in a DART recording prior to a tsunami, as a buoy construction feature \cite[]{dart1}, \cite[]{hal3}. In still conditions, a DART buoy transmits a series of 15-min data once every 6 hours. When a tsunami is detected or a buoy is triggered by an operator, the buoy switches to (nearly) real-time reporting of 1-min data. However, the 15-min data collected since the last transmission would be temporarily missing in the record until the standard reporting mode resumed.  

For most filtering methods, a signal in the immediate vicinity of a current point contributes to a current output with greater weight \cite[]{hamming}. Therefore, a gap in the data presents a serious difficulty in filtering the record. On the contrary, the EOF filter output at each point is computed with the entire input and can tolerate an absence of part of the data and eventually fill in data gaps (EOF filter interpretation as a bank of running windows does not apply anymore). 

In the presence of missing data, EOF coefficients ${\bf a}$ of a tidal fragment ${\bf y}$ can be found by minimizing the least-square error of fitting the existing data with EOFs:
\begin{equation}
F({\bf a})=\sum_{m=1}^M \alpha_m \left( {y(m)-\sum_{k=1}^n{a_k f_k(m)}} \right)^2 = \min
\label{aminim}
\end{equation}
where the $m$-th weight $\alpha_m$ equals 1 if the $m$-th element in ${\bf y}$ is present, and 0 otherwise. 
The similar approach was used in \cite[]{boyd} for interpolating missing data in ocean profiles and in \cite[]{gappy} to fill in gaps on images of a certain class, given a pre-computed EOF basis. 
Equation (\ref{aminim}) yields a set of linear equations with respect to ${\bf a}$:
\begin{equation}
gg^T \cdot gg \cdot {\bf a}=gg^T \cdot {\bf y}
\label{aequ}
\end{equation}
where 
$gg$ is a $M \times n$ matrix with elements ${g_i}(m)=\alpha_m f_i(m)$ in the $i$-th column. 
The system (\ref{aequ}) has a unique solution as long as the symmetrical $n \times n$ matrix $B=gg^T \cdot gg$ is not poorly conditioned. Then the EOF filter output is given by
\begin{equation}
{\bf z}= ff \cdot B^{-1} \cdot gg^T \cdot {\bf y}
\end{equation}
and yields a tidal component for the entire fragment. 

\begin{figure}[h]
	\resizebox{\textwidth}{!}
		{\includegraphics{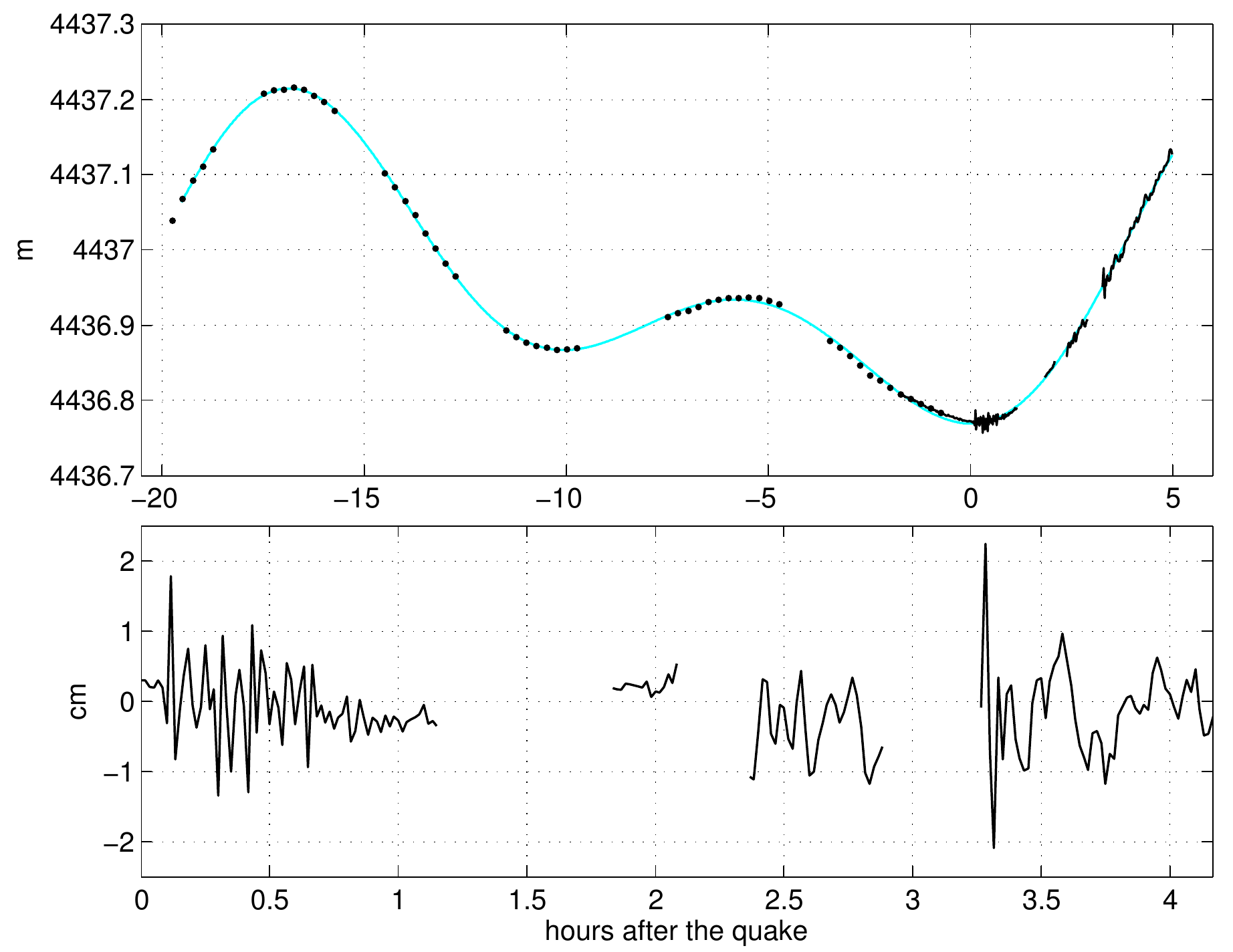}}
	\caption{
	Top pane: 01/03/2009 tsunami record at DART 52403 sampled at 15-min (black dots) and 1-min (black solid line) intervals, and the tidal component reconstructed via fitting with EOFs (cyan/gray line). Bottom pane: de-tided tsunami record.
	}
	\label{gappy}
\end{figure} 

Figure \ref{gappy} shows an example of de-tiding of an actual tsunami record with data gaps. The tsunami was recorded on 01/03/2009 at DART 52403. The record is a mixture of data sampled with 15-min (black dots) and 1-min (black solid line) intervals. Some of the data preceding the triggering might not be available at the time of the actual event.  

\begin{figure}[h]
	\centering
	\resizebox{0.9\textwidth}{!}
		{\includegraphics{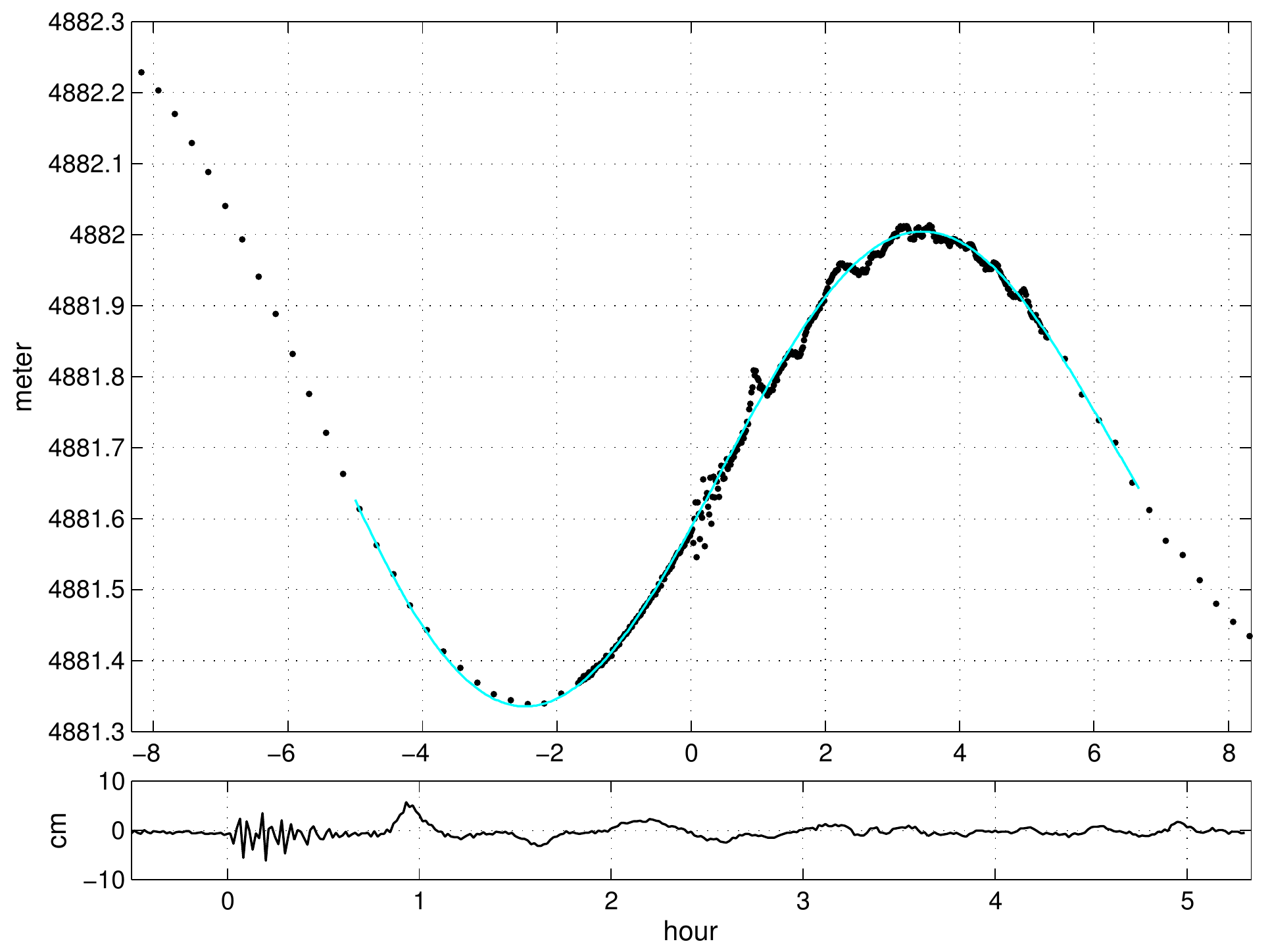}}
	\caption{
	Top pane: DART 32401 record of Peruvian tsunami of 08/2007 (black dots) and its tidal component according to a low-pass filter (gray/cyan). Bottom: de-tided tsunami record. Time is in hours since the quake.
	}
	\label{peru_ref}
\end{figure} 

\begin{figure}[h]
	\resizebox{\textwidth}{!}
		{\includegraphics{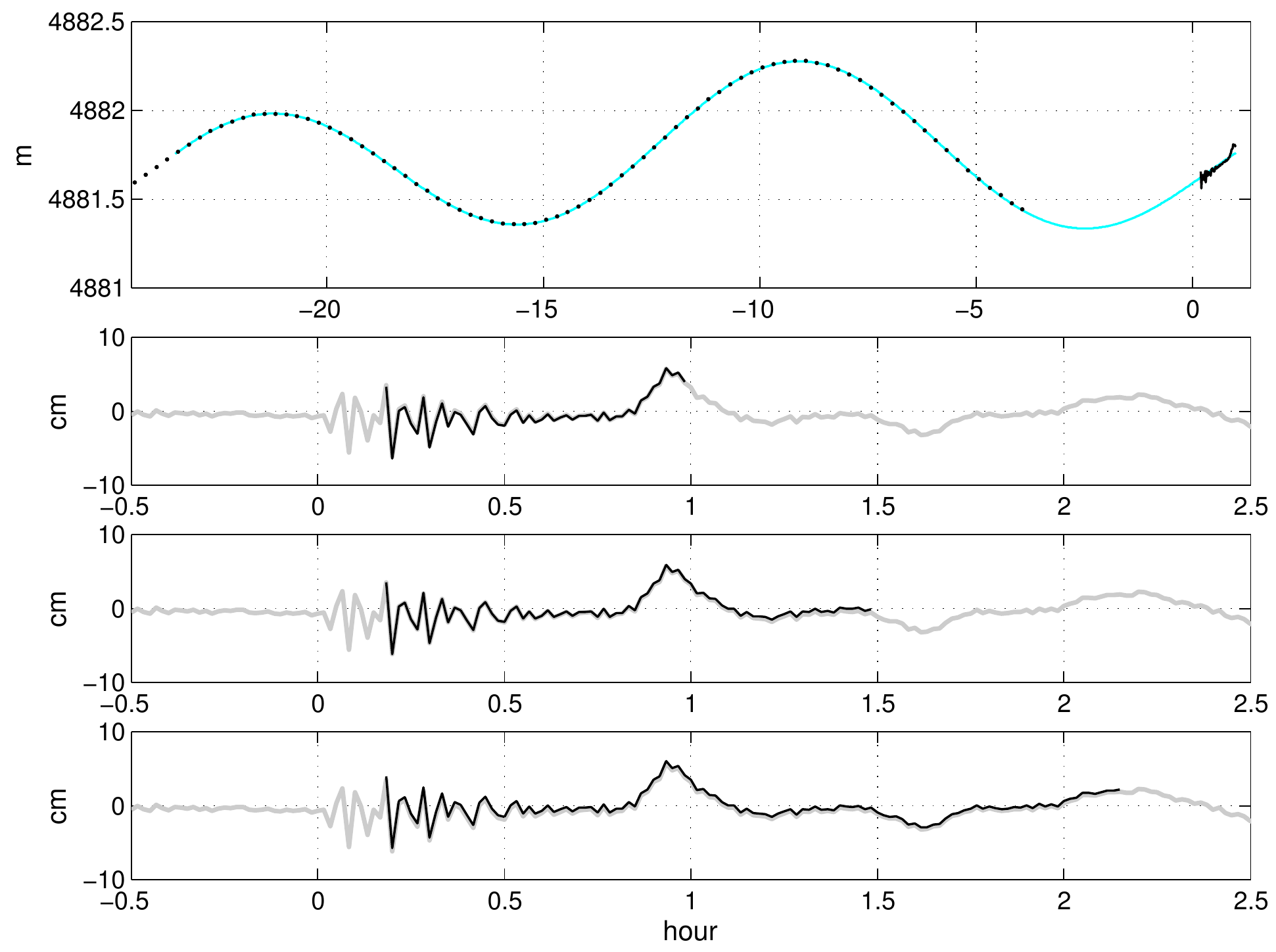}}
	\caption{
	Top pane: 32401 DART record imitating the ongoing Peruvian tsunami (black dots - 15-min data, black line - 1-min data) with a 4-hour gap preceding 1-min data, and an EOF filter output filling in the gap (cyan/gray). Bottom panes: reference tsunami signal (gray/thick) and EOF de-tided tsunami record at the DART (black/thin) starting at 11 min and ending at 60 min, 90 min, and 130 min after the quake.
	}
	\label{peru_eof}
\end{figure} 

The next example illustrates the EOF de-tiding "during an event". Figure \ref{peru_ref} shows a record of the Peruvian tsunami of August 2008 at DART 32401. The record contains data with 15-min (standard mode) and 1-min (event mode) sampling rates. One-min data prior to the earthquake (at $t=0$) were actually transmitted within an hour after the buoy entered an event mode \cite[]{dart1}. 
To extract a tidal component, the record was re-sampled to a 1-min rate and low-pass filtered using a 1-day wide cosine-Lanczos filter with 3.5 c/d cut-off. The difference between 1-min original data and the filter output, shown in the bottom pane in Fig. \ref{peru_ref}, is considered the reference tsunami signal. 
To imitate an ongoing event lasting time $\tau$, a 4-hour section of the data preceding the "event detection" (supposedly at the peak of the seismic signal) 
was taken out, and the entire record was cut at time $\tau$ mark (60 min, 90 min, and 130 min after the quake, or about 10 min, 40 min, and 70 min after the tsunami arrived at the buoy). 
Figure \ref{peru_eof}, top pane, shows a 1-lunar day long section of the record ending 60 min after the quake, processed with a 7+1 EOF filter. The processing yields a tidal component in the section and recovers the missing tidal data. After subtracting the tidal component, the tsunami signal is extracted. The bottom panes of the figure show a tsunami component of a record ending 60 min, 90 min, and 130 min after the quake, extracted via EOF de-tiding, against the reference tsunami signal. Very good agreement with the reference and no edge effects are observed.

\section{Predicting filter}
\begin{figure}[h]
\centering
	\resizebox{0.8\textwidth}{!}
		{\includegraphics{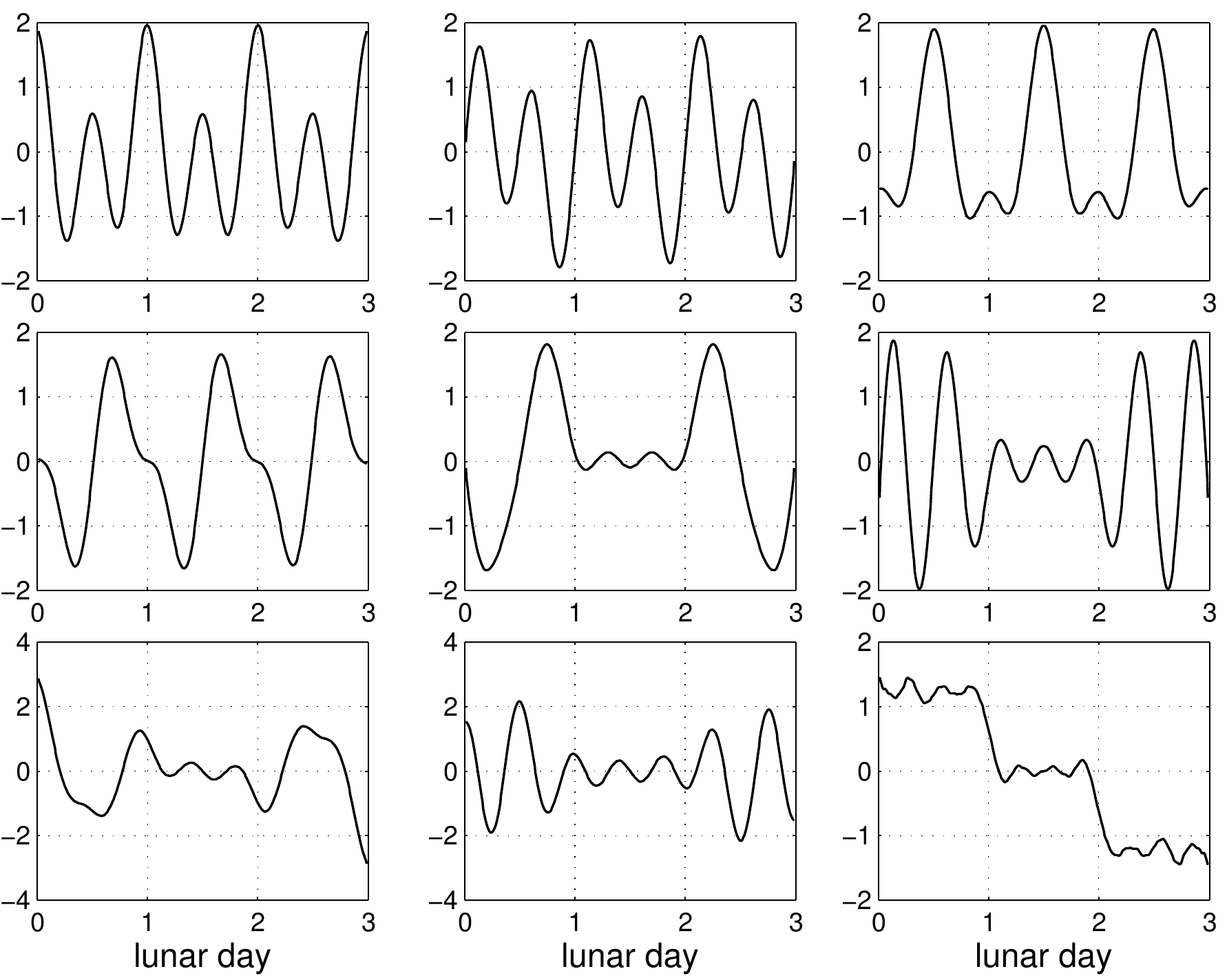}}
	\caption{
	The EOF set of 9 vectors (to be appended by a constant vector) for 3-day long tidal fragments. The EOF amplitude is shown in proportion to the constant vector amplitude.
	}
	\label{9eofs}
\end{figure} 

\begin{figure}[h]
\centering
	\resizebox{0.8\textwidth}{!}
		{\includegraphics{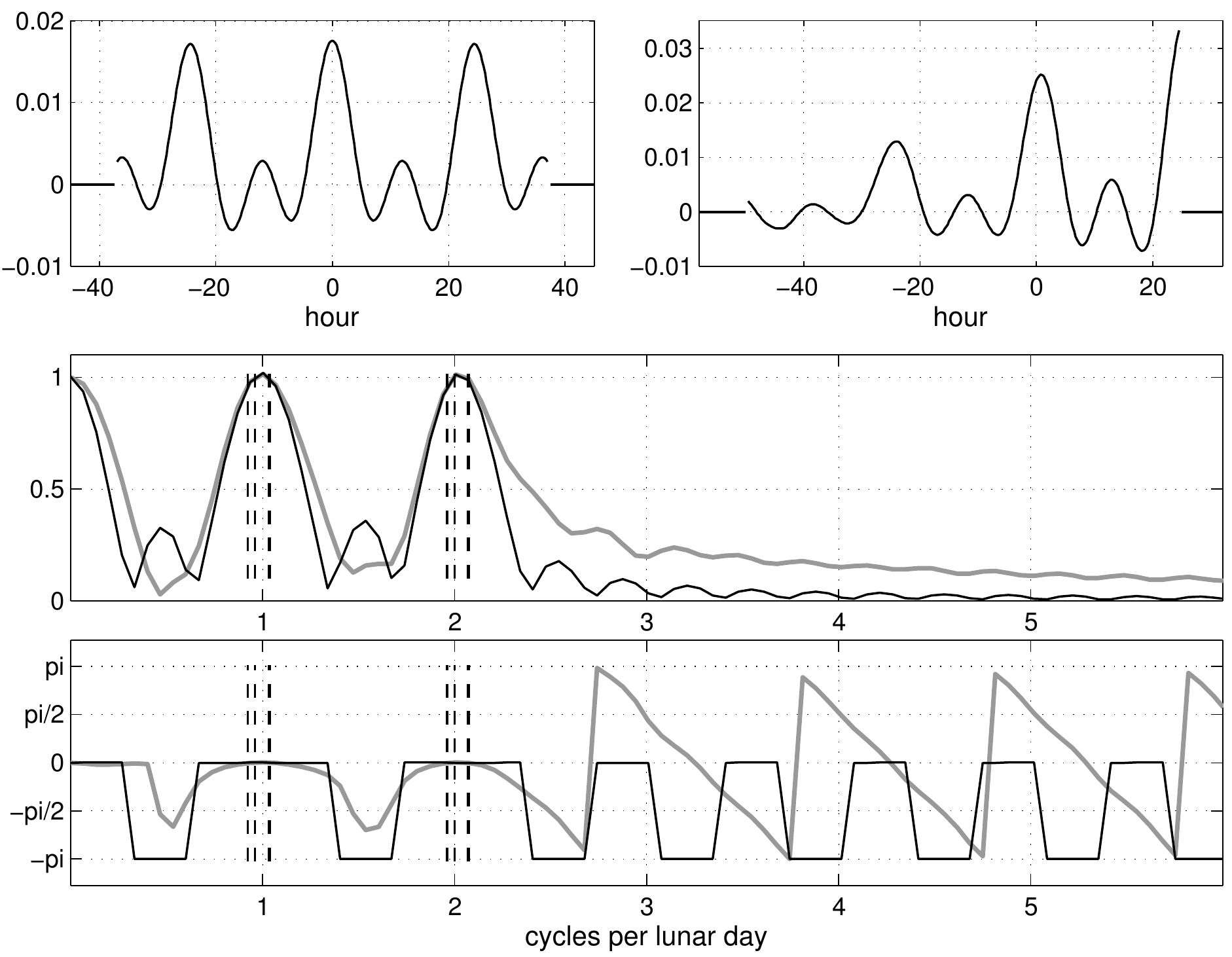}}
	\caption{
	Top panes, left to right: sliding windows in the central (149th) row of $297 \times 297$ matrix A and in 99th row. Middle pane: amplitude characteristics of the corresponding filters in the frequency domain (black/thin - 149th row, gray/thick - 99th row); bottom: phase characteristics of the filters. Dotted vertical lines denote the frequencies of the major eight harmonic constituents (listed in the order of increasing frequency):  $Q_1, O_1, P_1, K_1$ and $N_2, M_2, S_2, K_2$.
	}
	\label{A3}
\end{figure} 

Both diurnal and semidiurnal tidal bands have an effective bandwidth of 0.3 cycles per day. This suggests that an envelop of either oscillation (at 1 c/d and at 2 c/d) has a correlation interval of 3.3 days (3.2 lunar days). Originally, this was a reason why Munk and Cartwright (1966) chose to sample a tidal admittance function (a pulse response to the gravitational forcing) with a 1.7-day interval, rounded to 2 days. This also suggests an opportunity of short-term tidal predictions utilizing an inherent signal structure. 

\begin{figure}[h]
\centering
	\resizebox{0.8\textwidth}{!}
		{\includegraphics{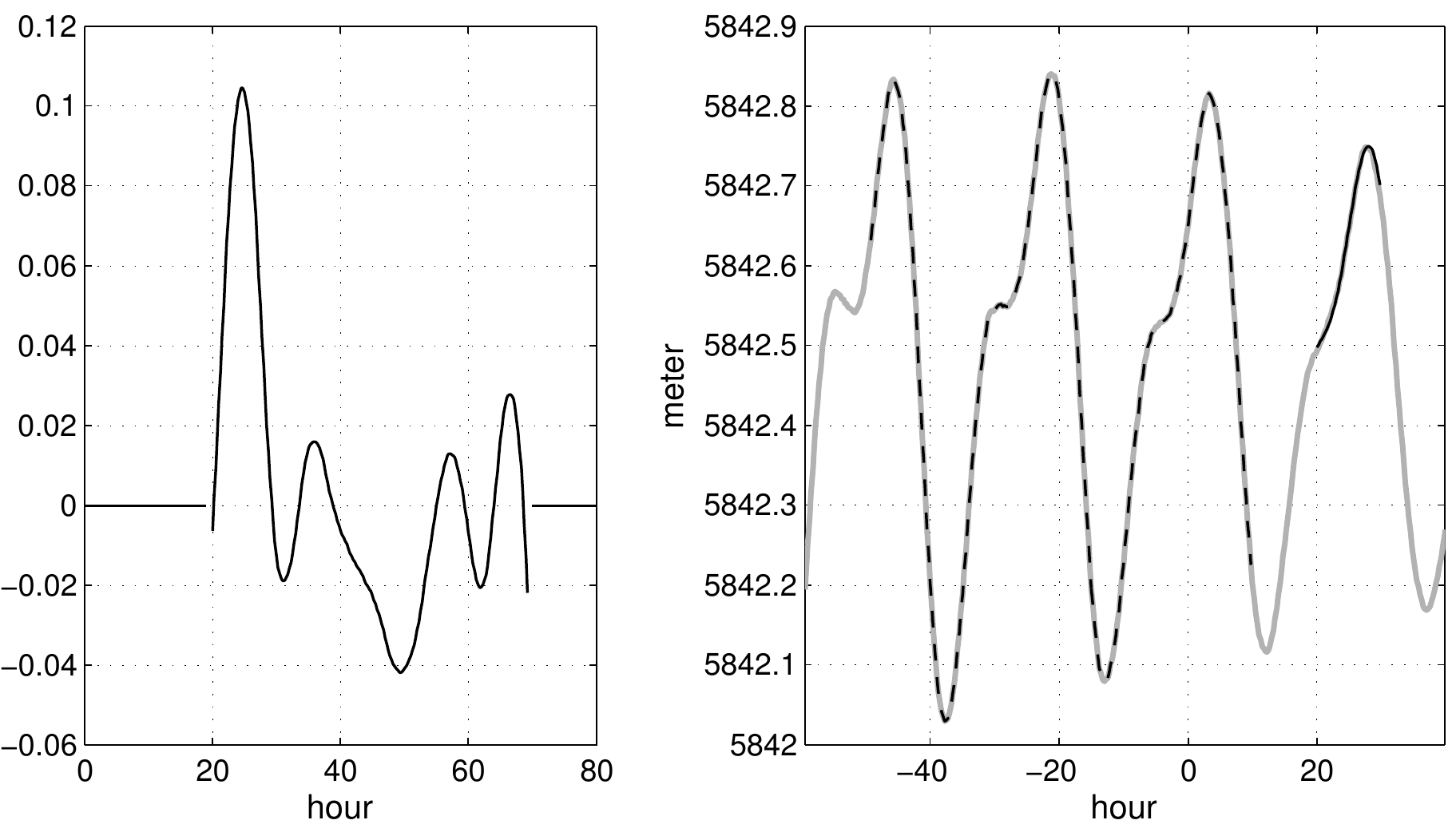}}
	\caption{
	Left: a predicting filter pulse response. Right: 10 hours of the filter output (black solid). Each output value requires 49.5 hours of input (black dashed) ending 20 hours prior to the output moment.  Thick gray line - tidal record of DART 21413.
	}
	\label{prd_pls}
\end{figure} 

\begin{figure}[h]
\centering
	\resizebox{0.8\textwidth}{!}
		{\includegraphics{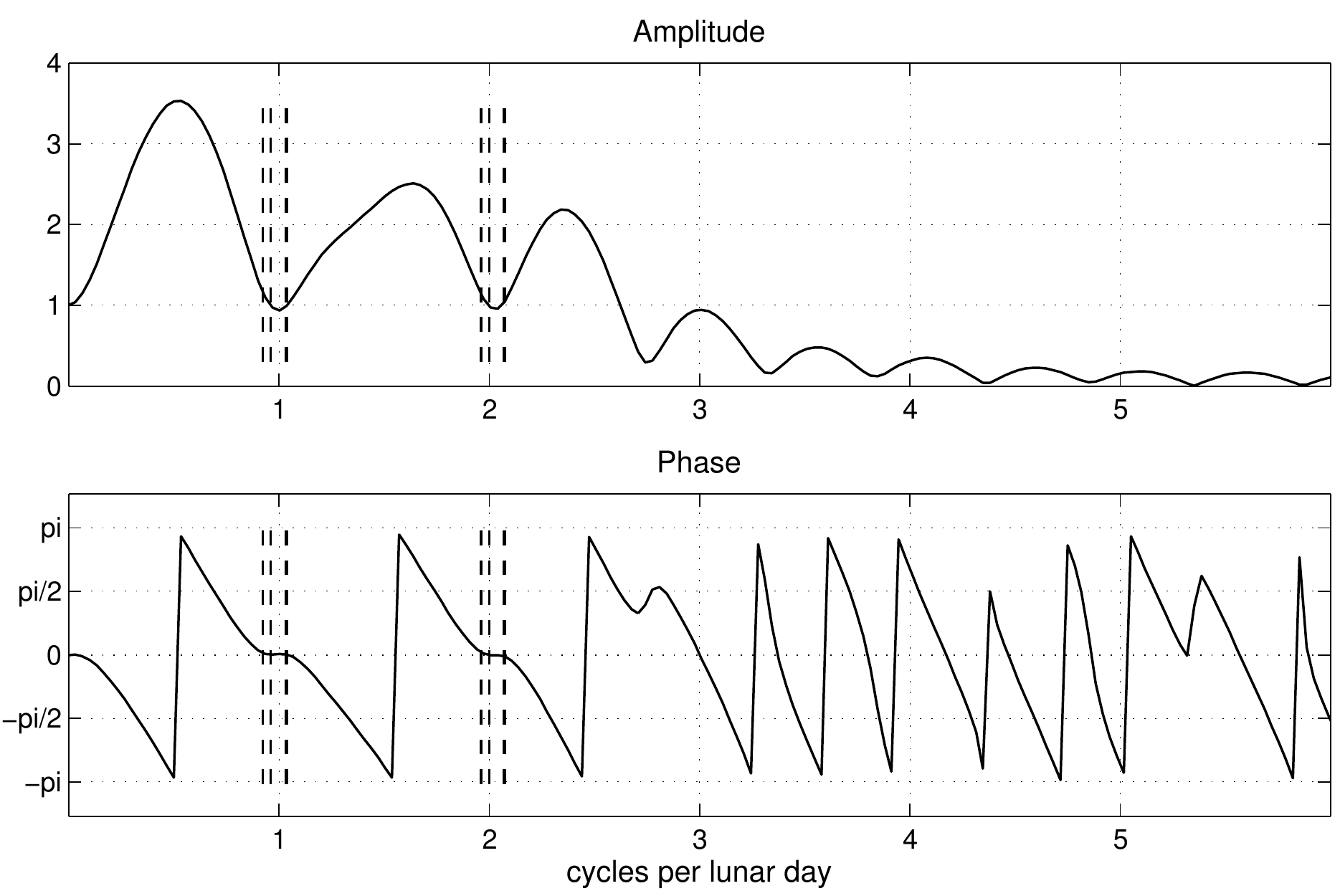}}
	\caption{
	Amplitude and phase characteristics of the filter predicting a tidal signal 20 hours ahead. Dotted vertical lines denote the frequencies of the eight major harmonic constituents (listed in the order of increasing frequency):  $Q_1, O_1, P_1, K_1$ and $N_2, M_2, S_2, K_2$.
	}
	\label{prd_chars}
\end{figure} 

\begin{figure}[h]
\centering
	\resizebox{0.9\textwidth}{!}
		{\includegraphics{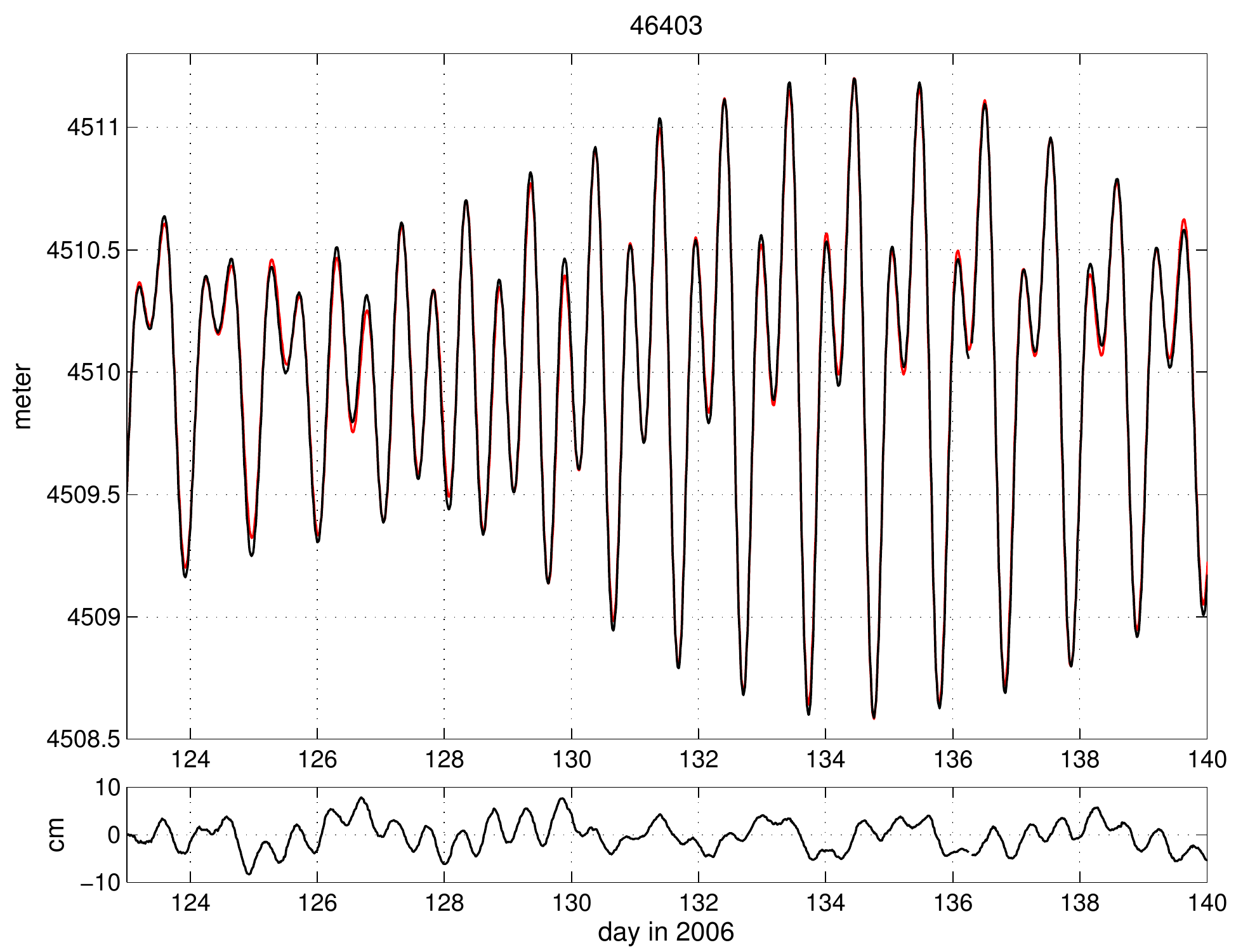}}
	\caption{
	Tidal record at DART 46403 (black), predicted tide (red), and prediction error (bottom pane). 
	}
	\label{pred46403}
\end{figure} 

\begin{figure}[h]
\centering
	\resizebox{0.9\textwidth}{!}
		{\includegraphics{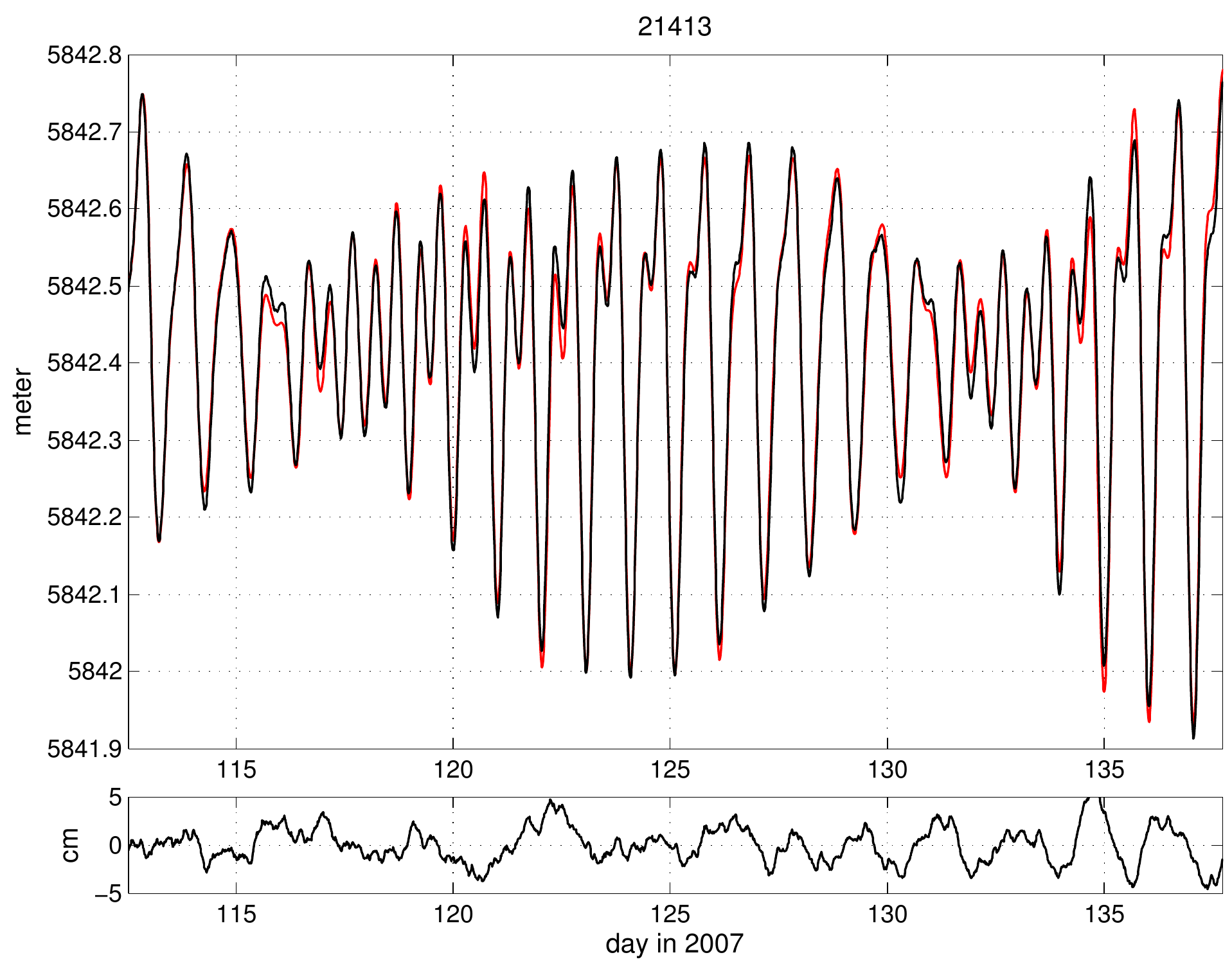}}
	\caption{
	Tidal record at DART 21413 (black), predicted tide (red), and prediction error (bottom pane). 
	}
	\label{pred21413}
\end{figure} 

\begin{figure}[h]
\centering
	\resizebox{\textwidth}{!}
		{\includegraphics{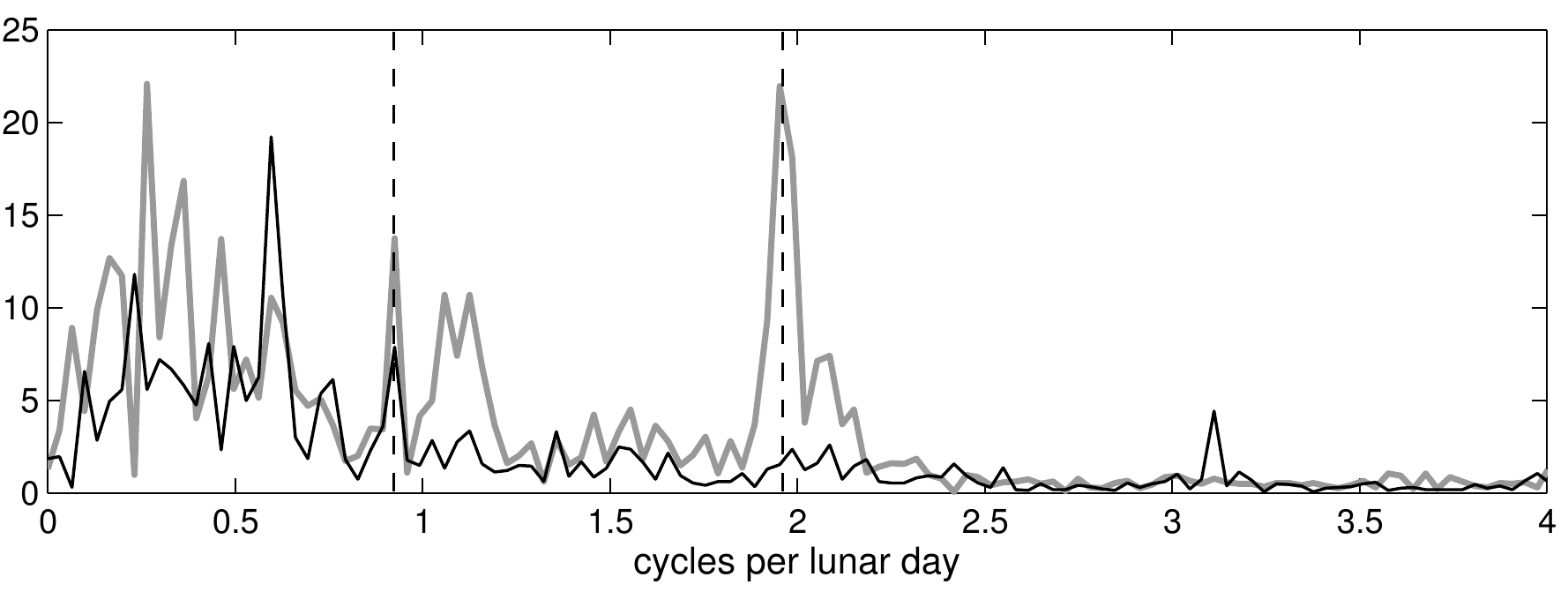}}
	\caption{
	Amplitudes of prediction error spectra at DART 21413 (black/thin) and 46403 (gray/thick). Dotted vertical lines denote the frequencies of $Q_1$ and $N_2$ primary tides.
	}
	\label{prderr}
\end{figure} 

The predictions are performed in a space of 3-lunar-day long tidal fragments. 
An EOF basis of 3-day long functions was developed following the same procedure, as with 1-day tidal fragments.
The tidal sub-space in $M=3 \cdot 99=297$ dimensional space, common for all the buoys, was found to include $n=10$ dimensions (9 EOFs with zero mean (shown in Fig. \ref{9eofs}) and a constant function). For this 9+1 basis set, a RMS error of approximation of 3-day long tidal fragments in the same records of 8 DARTs as in Fig. \ref{resds}, averaged among 30-60 fragments within the same record, varied from 0.6 cm to 1.2 cm.

Figure \ref{A3} shows pulse and frequency responses of the filters contained in the 99th and 149th (the central) row/column of a tidal analysis matrix A (\ref{mtrxA}) associated with the basis of 3-day long EOFs. The filters are transparent and introduce no phase distortions around the center of each tidal band and damp the signal components outside the three tidal bands, which explains the universality of the basis. However, the filter does not maintain a unit amplitude throughout each band, so it can not extract as much of the tidal variance as the 1-day EOF filter does. Consequently, the signal component orthogonal to the sub-space of the $n$  3-day long EOFs will be referred to as an EOF residual rather than a non-tidal component. 

Short-term tidal predictions can be viewed as filling in a data gap located past the last known reading. The prediction algorithm uses the latest record ${\bf y}_N$ of $N<M$ readings to predict $M-N$ future readings, with the first $N$ readings being used to estimate the EOF coefficients of the entire fragment of $M$ readings according to (\ref{aequ}), ${\bf y}$ being replaced with ${\bf y}_N$ and $gg$ being an $N \times n$ matrix with the first $N$ elements in vectors ${\bf f}_i$ for its columns. 
Let ${\bf b}$ denote a set of $n$ EOF coefficients of the entire fragment of $N$ recorded and $M-N$ future readings. Then
\begin{equation}
{\bf y}_N=gg \cdot {\bf b} + {\bf \eta}_N ,
\label{bequ}
\end{equation}
where ${\bf \eta}_N$ is the recorded part of an EOF residual in the fragment. 
If the residual were zero, then ${\bf a}= {\bf b}$, and the prediction was perfect. In a real situation, however, the EOF coefficients are found with an error
\begin{equation}
{\delta {\bf a}}=B^{-1} \cdot gg^T \cdot  {\bf {\eta}}_N ,
\label{deltaa}
\end{equation}
For $N \ge 198$ matrix $B$ is well-conditioned.
Thus, a 198 reading (49.5 hour) long record at any DART buoy can forecast the following 99 readings (the next lunar day record) at the same DART, simply by left-multiplying the record by a $(M-N) \times N$ matrix $P$ universal for all DARTs:
\begin{equation}
P= ee \cdot B^{-1} \cdot gg^T ,
\label{Pmatrix}
\end{equation}
where $ee$ is a $(M-N) \times n$ matrix with the last $M-N$ elements in vectors ${\bf f}_i$ for its columns. 

The underlying physics of the forecast can be revealed 
by following the same approach as described in section \ref{sec_filts} with respect to detiding. Matrix $P$ can be viewed as a bank of $M-N$ sliding windows of $N$ readings each. The $k$-th filter in the bank predicts a value to be recorded $k$ readings later than the last reading in the input.

In general, a filter to predict a tide should meet the following criteria:
\begin{itemize}
\item
the filter's amplitude characteristics is equal to 1 and the phase characteristics is equal to 0 within tidal bands; and
\item
the filter's pulse response is zero until some time $\tau>0$.
\end{itemize}
The first of the two conditions implies that if the signal at the filter input is a pure tide, then the signal at the output is the true tide as well. The second one ensures that the output is shifted by $\tau$ into the future with respect to the most recent input. Thus the output predicts a value to occur time $\tau$ later. \\

Figure \ref{prd_pls}, left pane shows a pulse response of a predicting filter in the 80th row of matrix $P$. The pulse is a finite function different from zero during 49.5 hours, the first non-zero reading occurring at the 20-hour marker. Thus 49.5 hours (198 readings) of recent tidal recording are used to predict a value to be recorded 20 hours (80 readings) later (Fig. \ref{prd_pls}, right pane). 
The amplitude and the phase of the filter frequency response are shown in Fig \ref{prd_chars}. The conditions for a predicting filter (1 for a filter amplitude and 0 for its phase), though not maintained band-wide, are met with the best accuracy exactly for the six major constituents $K_1, O_1, P_1$ and $M_2, S_2, K_2$. 
The forecast error should be attributed to an amplification of the other constituents within the tidal bands, first of all $Q_1$ and $N_2$, and to a non-tidal component (long wave noise especially), should it occur in a record. 
Noise amplification in tidal bands is not a shortcoming, but a necessary feature of a predicting filter. It ensures progressing degradation of predictions made recursively. Otherwise, a single 49.5 h long recording would define a tidal record indefinitely, which would be a violation of physical laws.

Figure \ref{pred46403} shows input and output of the above filter for a several day long record, and the difference of the two (prediction error). The input tide was a part of the 2006 record of DART 46403 with a mixed tide. The RMS prediction error per reading during a month long prediction was 3.0 cm (2.8 cm, if the predictions were done in 99 reading fragments obtained with the entire $P$), or less than 1\% of a tidal range ($>3$ m) at this DART. Figure \ref{pred21413} shows input and output for a several day long record, and the prediction error in a part of the 2007 record of DART 21413, which is dominated by the diurnal constituents. The RMS prediction error per reading during a month long prediction was 1.8 cm (1.7 cm, if the predictions were obtained with the entire matrix), or about 2\% of a tidal range (90 cm) at this DART. 

Amplitude spectra of the month-long prediction error for the two buoys are shown in Fig. \ref{prderr}. 
The major source of the prediction error is the long wave ocean noise in the range from 0 to 1 c/d, where the filter amplification is most significant. Also prominent in the error spectra are the constituents $Q_1$ and $N_2$ amplified by the filter as well.

Overall the predictions are at least as accurate as a harmonic-constituent-based forecast provides, with more EOF forecast examples and the accuracy comparisons in \cite[]{me}. 
\clearpage
\section{Conclusions}

As discussed in this work, fragments of deep-ocean tidal records up to 3 day long share the same functional sub-space, regardless of the record's origin.  The tidal sub-space basis can be derived via EOF analysis of a tidal record of a single buoy.
EOF analysis of a time series, however, assumes that the time series represents a stationary (in the weak sense) process.
In this work, a modification of one-dimensional EOF formalism is introduced that is not restricted to stationary processes. With this modification, 
the EOF-based de-tiding/forecasting technique is interpreted in terms of a signal passage through a filter bank, which is unique for the sub-space spanned by the EOFs. This interpretation helps to identify a harmonic content of a continuous process whose fragments are decomposed by given EOFs. In particular, 7 EOFs and a constant function are proved to decompose 1-day-long tidal fragments at any location. Filtering by projection into a reduced sub-space of the above EOFs is capable of isolating a tsunami wave within a few mm accuracy from the first minutes of the tsunami's appearance on a tsunami buoy record, and is reliable in the presence of data gaps.
EOFs with $\sim 3$ day duration (a reciprocal of either tidal band width) allow short-term (24.75 hours in advance) tidal predictions using an inherent structure of a tidal signal. The predictions do not require any a priori knowledge of tidal processes at a particular location, except for recent 49.5 hour long recording at the location. 

\section{Acknowledgments} 
This study and publication are funded by NOAA and by the Joint Institute for the Study of the Atmosphere and Ocean (JISAO) under NOAA Cooperative Agreement No. NA17RJ1232, Contribution 1606 (JISAO) and 3267 (PMEL). All the DART records used in this work have been obtained from NOAA's National Data Buoy Center public website \url{http://www.ndbc.noaa.gov/dart.shtml}. 
I thank my colleagues in the NOAA Center for Tsunami Research for their decision to implement an EOF filter as one of the de-tiding methods in the Short-term Inundation Forecasting for Tsunamis (SIFT) system.
Sincere thanks are due to Ryan L. Whitney for proofreading the manuscript.

\end{document}